\documentclass[12pt,aps,tightenlines]{revtex4-2}
\usepackage{amsmath}
\usepackage{amssymb}
\usepackage{bm}
\usepackage[mathscr]{eucal}
\usepackage{graphicx}
\usepackage{psfrag}
\usepackage{subfigure}
\usepackage{color}
\usepackage{wasysym}
\include{graphix}
\usepackage{framed}
\usepackage{enumitem}%
\usepackage{lipsum}
\usepackage{float}
\usepackage{stmaryrd}
\usepackage{MnSymbol}
\definecolor{orange}{rgb}{1,0.5,0}
\usepackage[normalem]{ulem}
\usepackage{cancel}
\usepackage{hyperref}

\usepackage{tikz}

\usetikzlibrary{patterns}
\usetikzlibrary{shapes}

\graphicspath{{figs/}}

\newtheorem{theorem}{Theorem}[section]
\newenvironment{proof}[1][Proof]{\begin{trivlist}
\item[\hskip \labelsep {\bfseries #1}]}{\end{trivlist}}

\newcommand{\myqed}{\nobreak \ifvmode \relax \else
      \ifdim\lastskip<1.5em \hskip-\lastskip
      \hskip1.5em plus0em minus0.5em \fi \nobreak
      \vrule height0.75em width0.5em depth0.25em\fi}

\newcommand{\mathd}{\mathrm{d}}
\newcommand{\mathe}{\mathrm{e}}

\newcommand{\vecu}{\bm{u}}

\newcommand{\hatz}{\widehat{\bm{z}}}
\newcommand{\hatn}{\widehat{\bm{n}}}
\newcommand{\hatt}{\widehat{\bm{t}}}
\newcommand{\hati}{\widehat{\bm{i}}}
\newcommand{\hatj}{\widehat{\bm{j}}}

\newcommand{\hatrho}{\widehat{\rho}}
\newcommand{\hatphi}{\widehat{\phi}}

\newcommand{\imag}{\mathrm{i}}

\newcommand{\myJ}{J}

\newcommand{\hatcL}{\widehat{c}}

\newcommand{\myInt}{\mathcal{I}}
\newcommand{\myJint}{\mathcal{J}}

\begin{document}

\title{Some analytical solutions for the evaporative flux and the flow inside a sessile droplet in cylindrical-cap geometry}

\author{Lennon \'O N\'araigh}
\email{Corresponding author.  Email: onaraigh@maths.ucd.ie}
\author{Conor Quigley}
\affiliation{School of Mathematics and Statistics, University College Dublin, Belfield, Dublin 4, Ireland}

\date{\today}

\begin{abstract}
We investigate the evaporation of sessile droplets in a two-dimensional (cylindrical-cap) geometry.  To circumvent the singularity in the fundamental solution of the Laplace problem for the vapour diffusion in two dimensions, previous works have imposed a far-field boundary condition at an arbitrary distance from the droplet.  To remove this arbitrary distance, we extend the approach of Yariv [Physical Review E, 107(6), p. L063101, 2023] for slender droplets (zero contact angles) to droplets with arbitrary contact angles.
Using conformal mapping, we obtain an explicit expression for the evaporative flux for arbitrary contact angles, and verify the result via numerical simulation.  We  couple the evaporative flux to the Stokes flow inside the droplet and use a local analysis to characterize the flow near the triple point. For contact angles below $90^\circ$, the evaporative flux exhibits the same power-law divergence in the two-dimensional geometry as in the corresponding three-dimensional (spherical-cap) geometry. The flow near the triple point likewise has the same power-law dependence in the two geometries. Our expressions for the evaporative flux and flow inside the droplet are explicit and involve no special functions, and therefore provide a simple explanation of the mechanism which drives the contact-line flow in evaporating sessile droplets.
\end{abstract}

\maketitle

\section{Introduction}
\label{sec:intro}

Droplet evaporation is an important physical phenomenon with applications in spray cooling, crop spraying, inkjet printing, and paint technology~\cite{wilson2023evaporation}.  In many applications, the focus is on sessile droplets.   This is a three-dimensional (albeit axisymmetric) phenomenon.   However, reduced two-dimensional models are common     in the literature~\cite{yarin2006lines,schofield2020shielding,yariv2023lifetime} as they greatly simplify the mathematics of droplet evaporation, and aid our understanding of same.  In this article we build up on these previous works, and present an analytical model which couples the vapour diffusion flux at the droplet surface to the flow inside the droplet, thereby providing a simple mechanistic description of the `coffee-ring effect' which is characteristic of droplet evaporation.   Before 
stating the aim of the paper
 in more concrete terms, we first place the work in the context of the literature.


References~\cite{yarin2006lines,schofield2020shielding,yariv2023lifetime} are the key works establishing the validity of the 2D modelling approach.  In References~\cite{yarin2006lines,schofield2020shielding}, a 2D model is used in which the sessile droplet takes the shape of a circular segment.  In this scenario, the vapour concentration $c$ in the atmosphere is determined by solving the steady-state diffusion equation:
\begin{equation}
\nabla^2c=0.
\label{eq:laplace}
\end{equation}
A standard Dirichlet boundary condition is applied on the free surface separating the droplet and the atmosphere (denoted by $S$):
\begin{subequations}
\begin{equation}
c=c_s,
\end{equation}
The fundamental solution of Equation~\eqref{eq:laplace} in 2D is $\ln r$, where $r=\sqrt{x^2+y^2}$ is the polar coordinate.  Because this fundamental solution is divergent as $r\rightarrow\infty$, the far-field boundary condition is imposed at a finite distance from the droplet:
\begin{equation}
c=c_\infty,\qquad r=R.
\end{equation}%
\label{eq:laplaceBC}%
\end{subequations}%
Finally, in that part of the atmosphere touching the substrate, a standard Neumann condition on $c$ applies.   A key quantity of interest arising from the solution of Equation~\eqref{eq:laplace} is the evaporative flux:
\begin{equation}
\myJ=-D_v \hatn\cdot\nabla c,\qquad \text{ on }S.
\label{eq:Jdef}
\end{equation}
where $D_v$ is the vapour diffusivity and $\hatn$ is the outward-pointing unit normal on $S$, pointing from the liquid into the atmosphere.

The conformal invariance of Laplace's equation in 2D is the main motivation for the 2D modelling approach.  Using conformal maps, Yarin \textit{et al.}~\cite{yarin2006lines} were able to show:
\begin{equation}
\myJ\sim \left(\frac{\ell}{a}\right)^{\lambda-1}
\end{equation}
where $\ell$ is the distance to the triple point measured in arc-length, and $2a$ is droplet footprint (equivalently, the chord length of the circular segment).  The exponent $\lambda$ is given by:
\begin{equation}
\lambda=\frac{\pi}{2(\pi-\alpha)},
\end{equation}
where $\alpha$ is the contact angle.  Thus, the flux diverges for sessile droplets, with $\alpha<\pi/2$.  The flux-divergence phenomenon (including the value of the exponent $\lambda$) are also well understood in the context of the full 3D model, as first computed by Popov~\cite{popov2005evaporative}.  The concordance between the 2D and 3D results highlights the usefulness in pursuing the 2D analytical modelling.

Reference~\cite{yarin2006lines} uses a conformal map from the circular segment to a half-disc, thereby simplifying the solution of the Laplace problem.  A similar approach is taken in Reference~\cite{schofield2020shielding}, where the conformal map is between an elliptical-shaped region (which approximates the circular segment), and a rectangular patch.  In the present approach, we use a M\"obius map which maps the physical space to a wedge.  A M\"obius map is useful here, because the droplet and the atmosphere are mapped to contiguous wedges, separated by a common ray.  This enables us to develop analytical solutions which couple together the conditions in the atmosphere and the droplet.   
In References~\cite{yarin2006lines,schofield2020shielding}, the  boundary condition at $r=R$ is problematic, since the value of the far-field boundary location $R$ is arbitrary.  And yet the far-field boundary condition $c=c_\infty$ cannot be imposed as $r\rightarrow\infty$ in 2D.    In a recent paper, Yariv~\cite{yariv2023lifetime} has resolved this paradox by recognizing that the correct model in 2D is in fact the time-dependent diffusion equation for the vapour concentration.  Under this assumption, Yariv has derived an expression for the evaporative flux of a cylindrical droplet in the small-contact-angle limit $\alpha\rightarrow 0$.  
 We summarize the features of the different 2D reduced models in Table~\ref{tab:summary}.


\begin{table}
	\centering
			\caption{Summary of the different models used to describe the 2D circular segment geometry}
		\begin{tabular}{|p{2cm}|p{1.8cm}|p{3cm}|p{1.8cm}|p{2.5cm}|}
		\hline
		Reference & Model & Conformal Map & Far-field BC & Main Result \\
		\hline
		\hline
	  Yarin \textit{et al.}~\cite{yarin2006lines} &  Laplace & Droplet $\leftrightarrow$ Half-Disc & At $r=R$ & Analytical flux expression for single droplet\\
		\hline
		Schofield \textit{et al.}~\cite{schofield2020shielding} &  Laplace & Ellipse-shaped region $\leftrightarrow$ Rectangular Patch & At $r=R$ & Analytical flux expression for droplet pair (shielding effect)\\
		\hline
		Yariv~\cite{yariv2023lifetime} &  Diffusion equation & N/A: Slender droplet, $\alpha=0$  & As $r\rightarrow\infty$ & Far-field boundary condition\\
		\hline
		This work &  Diffusion equation & Droplet $\leftrightarrow$ Wedge & As $r\rightarrow\infty$ &  Analytical expression for integral flux, coupling to flow inside droplet\\
		\hline
		\end{tabular}
\label{tab:summary}		
\end{table}

The point of departure in the present work is to marry all the separate approaches taken in References~\cite{yarin2006lines,schofield2020shielding,yariv2023lifetime}, and to develop a full time-dependent solution of the vapour-concentration equation, valid for all contact angles, with a far-field boundary condition imposed as $r\rightarrow\infty$.  To apply conformal-map techniques to the non-conformally-invariant diffusion equation, we use matched asymptotic expansions to stitch together an inner solution of the  Laplace's equation (which is conformally invariant) to an  outer solution of the diffusion equation (which is not conformally invariant, but is well characterized analytically).
 In doing this, we will uncover various useful expressions for the flux $\myJ$ as well as its integral,
\begin{equation}
F_{2D}=\int_S \myJ\mathd \ell,
\end{equation}
which may support our understanding of droplet evaporation.   Our expression for $F_{2D}$ links to the dynamics of droplet evaporation in both the constant-radius (CR) and constant contact-angle (CCA)  modes of evaporation~\cite{wilson2023evaporation}.


To understand the flow $\vecu$ inside the droplet, Stokes flow is assumed:
\begin{equation}
-\nabla p+\mu \nabla^2\vecu=0,\qquad \nabla\cdot\vecu=0,
\label{eq:stokes}
\end{equation}
where $p$ is pressure and $\mu$ is the dynamic viscosity.
Boundary conditions are: no-slip at the liquid-solid interface, and a stress-free condition on the surface $S$:
\begin{subequations}
\begin{equation}
\widehat{\bm{t}}\cdot\left(\nabla \vecu+\nabla\vecu^T\right)\cdot\hatn=0,\qquad\text{on }S.
\label{eq:noStress}
\end{equation}
Here, $\hatt$ is any unit tangent vector to the free surface $S$.
The velocity field $\vecu$ is coupled to the evaporation flux via the relation:
\begin{equation}
V_n+\frac{J}{\rho_\ell}=\hatn\cdot\vecu,\qquad\text{on }S,
\label{eq:massBC}
\end{equation}%
\label{eq:StokesBCs}%
\end{subequations}%
where $\rho_{\ell}$ is the density of the liquid inside the droplet.  Here, $V_n$ is the normal velocity of the surface $S$, which is non-zero for an evaporating droplet.  

Some solutions of Equation~\eqref{eq:StokesBCs} in the 3D geometry are given  in References~\cite{masoud2009analytical,lebedev2026stokes,wilson2023evaporation}.  Reference~\cite{masoud2009analytical} also studies the 2D case, and notes that Equation~\eqref{eq:laplace} does not have a solution in 2D when the outer boundary is at infinity, since the solution varies logarithmically.  The authors therefore study the Stokes problem~\eqref{eq:stokes} in 2D using a constant flux, for want of a better description.  Reference~\cite{gelderblom2012stokes} notes that in the vicinity of the triple point, the 3D spherical cap locally resembles a wedge, which can be described by the aforementioned 2D model~\eqref{eq:stokes}.   In this case, Reference~\cite{gelderblom2012stokes} solves Equation~\eqref{eq:stokes}  analytically to yield `corner eddies' similar to the ones described  by Moffatt~\cite{moffatt1964viscous}.  The different aspects of flow inside the evaporating droplet are outlined in detail in a recent review by Gelderblom \textit{et al.}~\cite{gelderblom2022evaporation}.

These disparate references provide important insights into a single unified physical phenomenon, namely the Deegan flow~\cite{deegan1997capillary} induced by the droplet evaporation.  For $\alpha<90^\circ$, droplet evaporation is particularly strong at the triple point, causing a compensatory flow inside the droplet, towards the contact line.  The effect is often referred to as the `coffee ring effect', since when a spilled drop of coffee dries on a solid surface, the Deegan flow prior to evaporation leaves a dense ring-like deposit along the droplet perimeter.


\subsection{Aims of the paper}

This paper builds on these previous works by bringing together a number of key strands, and has three main aims.  Firstly, the paper extends Yariv's~\cite{yariv2023lifetime} expression (2D diffusion equation, $\alpha=0$) for the evaporative flux  to arbitrary contact angles.   Next, the paper builds on the prior work by Yarin~\cite{yarin2006lines} by developing simple expressions for the evaporative flux and for the integral flux which are independent of an (arbitrary) far-field boundary location $R$.
Finally, the paper  extends the work of  Masoud and Felske~\cite{masoud2009analytical} by solving the Stokes-flow problem coupled to our physics-based expression for the 2D evaporative flux.

\subsection{Plan of the paper}

The paper is laid out as follows.  In Section~\ref{sec:theory} we introduce the mathematical model inside the atmosphere and inside the droplet.  We define the coordinate system and introduce the M\"obius map which simplifies the problem.  In Section~\ref{sec:atmosphere} we use the theory of matched asymptotic expansions to develop an approximate solution for the vapour concentration in the atmosphere, valid for arbitrary contact angles.  In Section~\ref{sec:numerics}, we verify the resulting expression for the integral flux via numerical simulation.  In Section~\ref{sec:droplet} we formulate the Stokes-flow problem in the mapped domain, and obtain asymptotic solutions valid in the vicinity of the triple point.  Discussion and concluding remarks are given in Section~\ref{sec:conc}.

\section{Theoretical Formulation}
\label{sec:theory}

We begin by fixing terminology.  Solving Equations~\eqref{eq:laplace}--\eqref{eq:stokes} in a two-dimensional ($xy$) planar geometry is equivalent to solving the same equations in a 3D planar ($xyz$) geometry with translational symmetry in the $z$ direction, corresponding to an infinite cylinder of liquid extending in the $z$-direction.  For this reason, throughout the paper, we refer to the 2D setup as a cylindrical droplet.  Naturally, such a configuration is not physically realistic, although the mathematical simplifications it produces enhance our understanding of the physics of droplet evaporation.

\begin{figure}[htb]
\centering
\begin{tikzpicture}[scale=1.2]

    \def\R{2.5}
    \def\d{1.0}
    \pgfmathsetmacro{\a}{sqrt(\R*\R-\d*\d)}
		\pgfmathsetmacro{\theta}{acos(\d/\R)}

		\fill[blue!20]
    (\a,\d)
    arc[start angle={90-\theta}, end angle={90+\theta}, radius=\R];
		
		\draw [blue,thick] (\a,\d) 
    (\a,\d)
    arc[start angle={90-\theta}, end angle={90+\theta}, radius=\R];
		
		\draw[thick,red,dashed] 
    (\a,\d) -- ({\a-0.8*\d}, {\d+0.8*\a});
		
		\draw
    ({\a-0.3*\d},\d)
    arc[start angle=180, end angle={180-atan(\a/\d)}, radius=0.3*\d]
    node[midway, above left] {$\alpha$};

    \draw[->] (3,3) -- (4,3) node[right] {$x$};
    \draw[->] (3,3) -- (3,4) node[above] {$y$};
		
		\draw[very thick,black] 
    (\a,\d) -- (5,\d);
		\node [draw,below] at (4,0.9) {$F_+$}; 
		\draw[very thick,green] 
    (-\a,\d) -- (-5,\d);
		\node [draw,below] at (-4,0.9) {$F_-$}; 
		
		\draw[very thick,green] 
    (-\a,\d) -- (-5,\d);
		\node [draw,below] at (0,0.9) {$F$}; 
		
    \draw[very thick,->] 
    ({\R*cos(90+25)},{\R*sin(90+25)}) --
    ({(\R+0.3)*cos(90+25)},{(\R+0.3)*sin(90+25)});
		\node [above] at ({(\R+0.25)*cos(90+25)},{(\R+0.25)*sin(90+25)}) {$\hatn$}; 

		\node [draw=blue,thick,align=center] at (0,3.5) {$\frac{\partial c}{\partial t}=D_v\nabla^2 c$\\$c=c_\infty$ for $t=0$};
		\node [draw=blue,thick,align=center,fill=white] at (0,1.7) {$-\nabla p+\mu \nabla^2\vecu=0$\\$\nabla\cdot\vecu=0$};
		
		
		\node[above]  at (-3.5,1.1*\d) {$\frac{\partial c}{\partial y}=0$};
		\node[right]  at (-2.6,2*\d) {$c=c_s$};
		\node[above]  at (-3.5,3*\d) {$\lim_{r\rightarrow\infty}c=c_\infty$};

		\draw[<->] (-\a,0.3*\d) --  (\a,0.3*\d);
		\node [below] at (0,0.2*\d) {$2a$};

		\node [draw,below] at (-2.4,1.8*\d) {$S$};
		
\end{tikzpicture}
\caption{Schematic description of the mathematical model.  The location of the origin of the coordinate system is left unspecified for the time being.  Domain boundaries are denoted as $F$, $F_-$, $F_+$, and $S$.}
\label{fig:schematic1}
\end{figure}
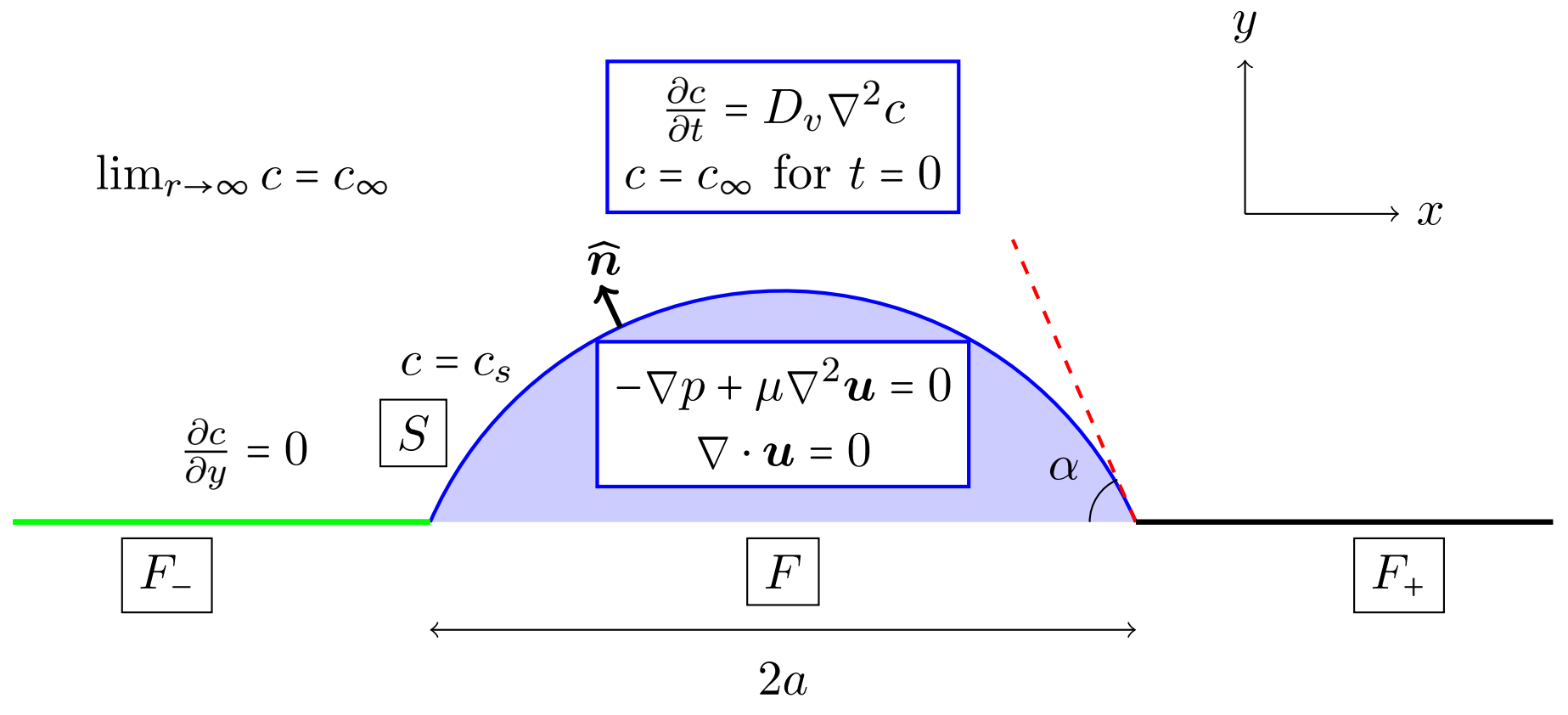

With this terminology in mind, a cylindrical droplet is placed on an infinite substrate as per Figure~\ref{fig:schematic1}.  The droplet base is $2a$ and the contact angle is $\alpha$.     Assuming that the droplet base  is small compared to the capillary lengthscale, capillarity dominates the shape of the free surface, which therefore assumes the shape of a circular segment (the 3D analogue is a spherical cap).  
The droplet is exposed to a moist atmosphere, and we assume the evaporation is diffusion-limited, and characterized by a uniform vapour diffusivity $D_v$.  In diffusion-limited evaporation, 
 the flow inside the droplet (domain: $\Omega_d$) decouples from the diffusion of vapour inside the atmosphere (domain: $\Omega_a$).  Consequently, the physical model in the atmosphere decouples completely from that inside the droplet~\cite{wilson2023evaporation}.  We use the following subsections to describe in more detail the mathematical models that apply in the separate domains.

\subsection{The model in the atmosphere}

The transport of liquid vapour inside the atmosphere is assumed to be diffusion-dominated~\cite{popov2005evaporative}.  The vapour concentration is therefore described by the diffusion equation:
\begin{subequations}
\begin{equation}
\frac{\partial c}{\partial t}=D_v\nabla^2 c,\qquad \text{in }\Omega_a,\qquad t>0.
\end{equation}
At the interface between the substrate and the atmosphere (faces $F_-$ and $F_+$), Neumann boundary conditions apply:
\begin{equation}
\frac{\partial c}{\partial y}=0,\qquad \text{on }F_+\text{ and }F_-.
\end{equation}
On the free surface $S$ separating the droplet and the atmosphere, the vapour concentration is taken as saturated:
\begin{equation}
c=c_s\qquad \text{on }S.
\end{equation}
Finally, the far-field boundary condition is:
\begin{equation}
\lim_{r\rightarrow \infty}c=c_\infty.
\end{equation}
Here, we have introduced the radial coordinate in the $xy$ plane, %
\begin{equation}
r=\sqrt{x^2+y^2}.
\end{equation}
Since the fundamental solution of $\nabla^2 c=0$ is proportional to $\ln r$, the two-dimensional diffusion equation with the foregoing boundary conditions does not admit a steady-state solution.  Consequently, the full time-independent diffusion equation must be solved and for that reason, an initial condition is required, which we take here to be:
\begin{equation}
c=c_\infty,\qquad t=0.
\end{equation}%
\label{eq:model_all}%
\end{subequations}%
The evaporative flux from the droplet into the atmosphere is given by Equation~\eqref{eq:Jdef} and drives the flow inside the droplet, as we now describe.

\subsection{The model inside the droplet}

The flow inside the droplet is characterized by its low Reynolds number~\cite{oron1997long} and consequently, may be described by the previously-introduced Stokes equation~\eqref{eq:stokes}.  The flow field $\vecu$ is assumed to be incompressible, $\nabla\cdot\vecu=0$.  Consequently, $\vecu$ may be described in terms of a streamfunction:
\begin{equation}
u=\frac{\partial \psi}{\partial y},\qquad v=-\frac{\partial\psi}{\partial x}.
\end{equation}
Under these conditions,  Equation~\eqref{eq:stokes} reduces to the biharmonic equation:
\begin{equation}
\nabla^4\psi=0,\qquad \text{in }\Omega_d.
\label{eq:biharmonic}
\end{equation}
The boundary conditions on $\vecu$ are no-slip on the substrate face $F$, which equates to:
\begin{equation}
\psi_x=0,\qquad \psi_y=0,\qquad \text{on }F,
\end{equation}
as well as the previously-introduced stress-free boundary condition~\eqref{eq:noStress}.
 To close the system of equations, a final boundary condition is required.  This is obtained by imposing the conservation of mass across the free surface $S$, the precise statement of which is given in Equation~\eqref{eq:massBC}.
%
%
%
An exact expression for $V_n$ in Equation~\eqref{eq:massBC} can be obtained in either the constant-radius (CR) or constant-contact-angle (CA) modes of evaporation.

Since we are dealing with a fundamentally two-dimensional problem, some of the physical quantities have non-standard units.  Therefore, to clarify the units used in this paper, a summary of physical quantities and their units is provided in Table~\ref{tab:units}.
\begin{table}[htb]
	\centering
		\begin{tabular}{|p{2cm}|p{3cm}|p{2cm}|}
		\hline
		Symbol & Quantity & Physical Units \\
		\hline
		\hline
			$c$ & Concentration & $\mathrm{M}/\mathrm{L}^2$\\
			$D_v$ & Diffusion coefficient & $\mathrm{L}^2/\mathrm{T}$\\
			$Q$   & Source term & $\mathrm{M}/\mathrm{T}$\\
			$\myJ$ & Evaporative flux & $\mathrm{M}/\left(\mathrm{L}\cdot\mathrm{T}\right)$ \\
			$F_{2D}$ & Integral flux in 2D & $\mathrm{M}/\mathrm{T}$\\
			$\rho_\ell$ & Mass per unit area (2D density) & $\mathrm{M}/\mathrm{L}^2$\\
			\hline
		\end{tabular}
		\caption{List of key quantities in the mathematical model and their physical units}
\label{tab:units}
\end{table}

\subsection{Coordinate System and M\"obius Transformation}

To simplify the foregoing mathematical analysis, the droplet base is offset from the horizontal axis of the coordinate system as shown in Figure~\ref{fig:schematic2}. 
The atmosphere $\Omega_a$ can therefore be described in mathematical terms as:
\begin{equation}
\Omega_a=\{(x,y)\in \mathbb{R}^2| x^2+y^2>R,\,y>d\}.
\end{equation}
The triple points where the atmosphere, substrate and liquid all meet are located at $y=d$ and $x=\pm a$, where $a=\sqrt{R^2-d^2}$.  The droplet base $a$, the contact angle $\alpha$, and the value $d$, and the radius of the circle are related via:
\begin{equation}
R\cos\alpha=d,\qquad R\sin\alpha=a.
\end{equation}
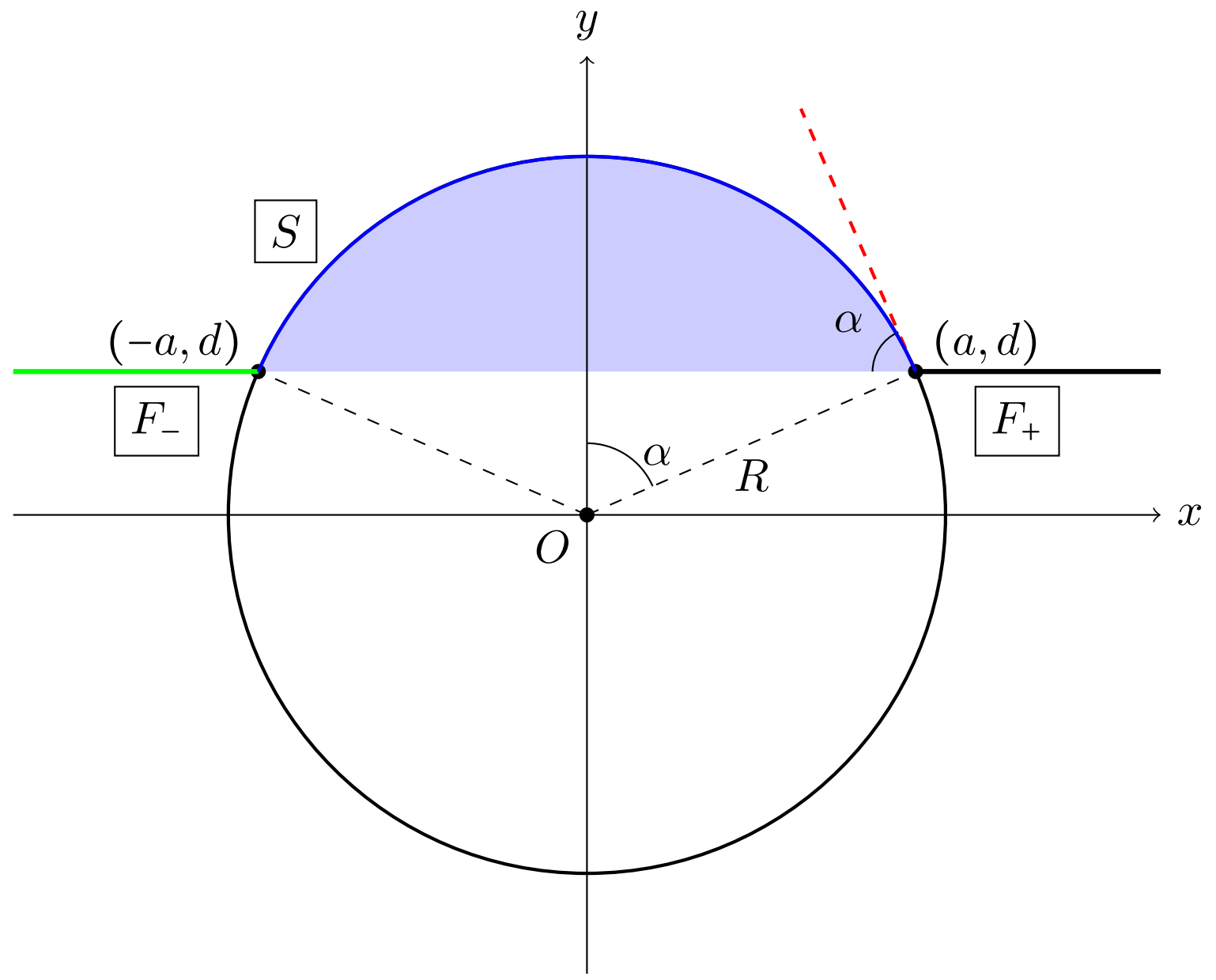
\begin{figure}
\centering
\begin{tikzpicture}[scale=1.2]

    \def\R{2.5}
    \def\d{1.0}
    \pgfmathsetmacro{\a}{sqrt(\R*\R-\d*\d)}
		\pgfmathsetmacro{\theta}{acos(\d/\R)}

\fill[blue!20]
    (\a,\d)
    arc[start angle={90-\theta}, end angle={90+\theta}, radius=\R]
    -- cycle;

    \draw[->] (-4,0) -- (4,0) node[right] {$x$};
    \draw[->] (0,-3.2) -- (0,3.2) node[above] {$y$};

    \draw[thick] (0,0) circle (\R);


    \fill (-\a,\d) circle (1.5pt);
		\node[left] at (-\a,1.2*\d) {$(-a,d)$};
    \fill (\a,\d) circle (1.5pt);
		\node[right] at (\a,1.2*\d) {$(a,d)$};

    \draw[dashed] (0,0) -- (\a,\d);
		\node[below] at (0.5*\a,0.5*\d) {$R$};
		
		\draw[dashed] (0,0) -- (-\a,\d);

    \fill (0,0) circle (1.5pt);
    \node[below left] at (0,0) {$O$};
		
		\draw (0,0.5) 
    arc[start angle=90, end angle={90-\theta}, radius=0.5] 
    node[midway, right] {$\alpha$};
		
		\draw[thick,red,dashed] 
    (\a,\d) -- ({\a-0.8*\d}, {\d+0.8*\a});
		
		\draw
    ({\a-0.3*\d},\d)
    arc[start angle=180, end angle={180-atan(\a/\d)}, radius=0.3*\d]
    node[midway, above left] {$\alpha$};
		
		\draw[very thick,black] 
    (\a,\d) -- (4,\d);
		\node [draw,below] at (3,0.9) {$F_+$}; 
		\draw[very thick,green] 
    (-\a,\d) -- (-4,\d);
		\node [draw,below] at (-3,0.9) {$F_-$}; 
		
		\draw [blue,thick] (\a,\d) 
    (\a,\d)
    arc[start angle={90-\theta}, end angle={90+\theta}, radius=\R];
		\node [draw,below] at (-2.1,2.2*\d) {$S$}; 
\end{tikzpicture}%
\caption{Schematic description of the chosen coordinate system}%
\label{fig:schematic2}%
\end{figure}%
To conformally map the domains in Figure~\ref{fig:schematic1} into a simpler ones, we introduce a complex coordinate $z=x+\imag y$.  The triple points are given by:
\begin{equation}
z_-=-a+\imag d,\qquad z_+=a+\imag d,
\end{equation}
or, in terms of polar coordinates,
\begin{equation}
z_-=R\mathe^{\imag \theta_-},\qquad z_+=R\mathe^{\imag\theta_+},
\end{equation}
where
\begin{equation}
\theta_-=\pi/2+\alpha,\qquad \theta_+=\pi/2-\alpha.
\label{eq:thetaPM}
\end{equation}
We can also write:
\begin{equation}
z_+-z_-=2R\sin\alpha.
\label{eq:zpmDiff}
\end{equation}
We introduce a the M\"obius transformation:
\begin{equation}
\zeta=f(z),\qquad f(z)=-\mathe^{\imag\alpha}\left(\frac{z-z_+}{z-z_-}\right).
\label{eq:Mobius}
\end{equation}
We write:
\begin{equation}
\zeta=\xi+\imag\eta.
\end{equation}
The inverse transformation is given by:
\begin{equation}
z=f^{-1}(\zeta),\qquad f^{-1}(\zeta)=\frac{\zeta z_-+\mathe^{\imag\alpha}z_+}{\zeta+\mathe^{\imag\alpha}}.
\label{eq:inverseT}
\end{equation}
The derivative is given by:
\begin{equation}
\frac{\mathd f}{\mathd z}=-\mathe^{\imag\alpha}\frac{z_+-z_-}{(z-z_-)^2}.
\label{eq:dfdz}%
\end{equation}
The derivative of the inverse is given by:
\begin{equation}
\frac{\mathd f^{-1}}{\mathd \zeta}=-\mathe^{\imag\alpha}\frac{z_+-z_-}{(\zeta-\mathe^{\imag\alpha})^2}.
\label{eq:dfInvdzeta}%
\end{equation}
In this work, it will be convenient to work with polar coordinates in the mapped space, with $\rho=\sqrt{\xi^2+\eta^2}$ and $\phi$ the polar angle starting from the $\xi$-axis and in tending in the direction of  increasing $\eta$.  In this context, we use the identity:
\begin{equation}
\frac{\mathd f^{-1}}{\mathd \zeta}\frac{\mathd f}{\mathd z}=1
\end{equation}
together with Equation~\eqref{eq:dfInvdzeta} to write:
\begin{equation}
\left|\frac{\mathd f}{\mathd z}\right|=\frac{\rho^2+2\rho\cos(\phi-\alpha)+1}{2a}.
\label{eq:dfdzMetric}
\end{equation}
Equation~\eqref{eq:dfdzMetric} will be particularly useful as it represents a metric factor used to transform between different orthogonal coordinate systems.

\begin{figure}
\centering
\begin{tikzpicture}[scale=1.1]

    \def\L{5.0}
    \def\myalpha{30}

		\fill[
    pattern=north east lines,
    pattern color=blue!50
    ]
    (0,0)
    -- ({\L*cos(180+\myalpha)}, {\L*sin(180+\myalpha)})
    arc[start angle={180+\myalpha}, end angle=360, radius=\L]
    -- cycle;
		
    \fill[blue!20]
    (0,0)
    -- ({\L*cos(\myalpha)}, {\L*sin(\myalpha)})
    arc[start angle={\myalpha}, end angle=0, radius=\L]
    -- cycle;
		
		\node [draw,right,fill=white] at (2.5,-2.5) {$\widetilde{\Omega_a}$}; 
		\node [draw,right,fill=white] at (2.5,1) {$\widetilde{\Omega_d}$}; 

    \draw[->] (-1.0,0) -- (1.2*\L,0) node[right] {$\xi$};
    \draw[->] (0,-1) -- (0,2) node[above] {$\eta$};
		\draw (0,-1.1*\L) -- (0,-1.5);

    \draw[very thick,blue,->] (0,0) -- (1.1*\L,0);
		\node  [draw,below,fill=white]  at (2.5,-0.1) {$\widetilde{S}$}; 

    \draw[very thick,green]
        (0,0) -- ({1.5*cos(180+\myalpha)}, {1.5*sin(180+\myalpha)});
		\node [draw,above] at (-1.2,-0.55) {$\widetilde{F_+}$}; 

	  \draw[very thick,black,->]
		   ({1.5*cos(180+\myalpha)},{1.5*sin(180+\myalpha)})--({1.1*\L*cos(180+\myalpha)},{1.1*\L*sin(180+\myalpha)});
		\node [draw,above] at (-3,-1.5) {$\widetilde{F_-}$}; 
		
		\draw[very thick,black,->]
		   (0,0)--({1.1*\L*cos(\myalpha)},{1.1*\L*sin(\myalpha)});
		\node [draw,above] at (3,2) {$\widetilde{F}$}; 


		
		\fill[red] (0,0) circle (0.1cm);
		\node[
		above right,
    align=center,
    fill=white,
    inner sep=2pt] at (0.15,0.15) 
		{
		{\textcolor{red}{$f(z_+)=0$}}
		};

    \node[above right] at (-2,1.5) {$\zeta=\xi+i\eta$};
		
		\fill[red] ({1.5*cos(180+\myalpha)}, {1.5*sin(180+\myalpha)}) circle (0.1cm);
		\node[
		above right,
    align=center,
    fill=white,
    inner sep=2pt] at (-1.6,-1.8) 
		{
		{\textcolor{red}{$\displaystyle\lim_{\substack{|z|\to \infty\\ z\in F_-\cup F_+}}f(z)
      =-\mathe^{\imag\alpha}$}}
		};
		
		\node[above right,
    align=center,
    fill=white,
    inner sep=2pt] at ({\L*cos(180+\myalpha)},{\L*sin(180+\myalpha)-0.7}) 
		{{\textcolor{red}{$\displaystyle\lim_{\substack{z\to z_-\\z\in F}}f(z)
      =-\infty e^{i\alpha}$}}};
		
		\node[above right,
    align=center,
    fill=white,
    inner sep=2pt] at (4,-0.8) 
		{{\textcolor{red}{$\displaystyle\lim_{\substack{z\to z_-\\z\in S}}f(z)
      =+\infty$}}};

\end{tikzpicture}
\caption{Schematic diagram showing the mapped domains after application of the M\"obius transformation.    
The wedges $\widetilde{\Omega_a}$ and $\widetilde{\Omega_d}$ are infinite in extent.  The hatched and filled areas in the figure are subsets of these infinite wedges and are shown to guide the eye.}
\label{fig:mappedDomain}
\end{figure}
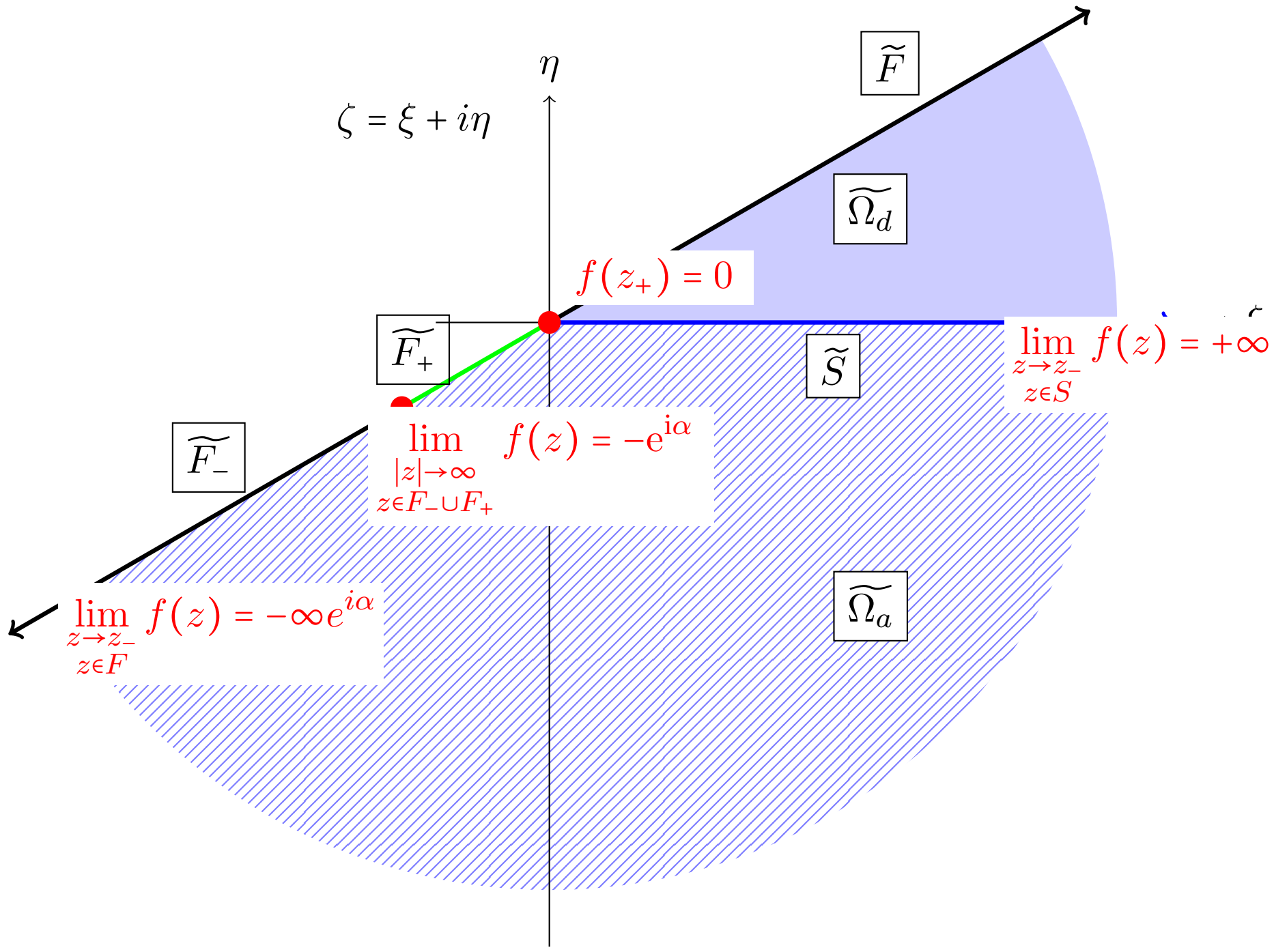

Under the M\"obius transformation, the domains $\Omega_a$ and $\Omega_d$  are mapped to contiguous wedges, as shown in Figure~\ref{fig:mappedDomain}.  We use a tilde to denote mapped domains and mapped faces 
(e.g. $\widetilde{\Omega_a}=f(\Omega_a)$, etc.).  For the avoidance of doubt, we enumerate the mapped domains and mapped boundaries as follows.
\begin{enumerate}[noitemsep]
\item Atmospheric domain $\Omega_a$: Mapped to the wedge $\widetilde{\Omega_a}$, with $\pi+\alpha<\arg(\zeta)<2\pi$.
\item Droplet domain $\Omega_d$: Mapped to the wedge $\widetilde{\Omega_d}$, with $0<\arg(\zeta)<\alpha$.
\item Boundary face $F_-$: Mapped to the interval $\zeta\in(-\infty,-1)\mathe^{\imag\alpha}$;
     \begin{enumerate}[noitemsep]
		 \item The triple point $z_-$ gets mapped to $-\infty\mathe^{\imag\alpha}$,
		 \item The far-field point $x\rightarrow-\infty$ gets mapped to $-\mathe^{\imag\alpha}$.
		\end{enumerate}
\item Boundary face $F_+$: Mapped to the interval $\zeta\in (-1,0)\mathe^{\imag\alpha}$;
     \begin{enumerate}[noitemsep]
		 \item The triple point $z_+$ gets mapped to $-\mathe^{\imag\alpha}$.
		 \end{enumerate}
\item Boundary face $F$: Mapped to the ray $\zeta\in (0,\infty)\mathe^{\imag\alpha}$.
\item Free surface $S$: Mapped to the ray $\zeta\in(0,\infty)$.
\end{enumerate}
Detailed calculations underpin these transformations and are provided in the Supplementary Material.   An important outcome of these calculations is the parametrization of the free surface $S$ in terms of polar coordinates in the $xy$ coordinate system.  On $S$, we have 
\begin{equation}
z=R\mathe^{\imag\theta},\qquad \theta_+<\theta<\theta_-,
\label{eq:thetaPlusDef}
\end{equation}
where $\theta_{\pm}$ are as defined previously in Equation~\eqref{eq:thetaPM}.  The $z$-values in Equation~\eqref{eq:thetaPlusDef} get mapped to:
\begin{equation}
\zeta=-\mathe^{\imag\alpha} \left(\frac{R\mathe^{\imag\theta}-R\mathe^{\imag\theta_+}}{R\mathe^{\imag\theta}-R\mathe^{\imag \theta_-}}\right).
\end{equation}
Using trigonometric identities, this can be re-written as:
\begin{equation}
\zeta=\xi+\imag\eta,\qquad \xi=-\frac{\sin\left(\frac{\theta-\theta_+}{2}\right)}{\sin\left(\frac{\theta-\theta_-}{2}\right)},\qquad \eta=0.
\label{eq:rhoTheta}
\end{equation}


We summarize the atmospheric model in the mapped space $\widetilde{\Omega_a}$.  The mapped diffusion equation reads:
\begin{subequations}
\begin{equation}
\frac{\partial c}{\partial t}=D_v|f'(z)|^2\nabla_{\xi\eta}^2c,\qquad \text{in } \widetilde{\Omega_a},\qquad t>0,
\label{eq:diffusionMapped}
\end{equation}
where $\nabla_{\xi\eta}^2=\partial_{\xi\xi}+\partial_{\eta\eta}$.
The boundary conditions ($t>0$)  on the mapped faces are:
\begin{eqnarray}
c&=&c_\infty,\qquad t=0,\\
-\sin\alpha\frac{\partial c}{\partial\xi}+\cos\alpha\frac{\partial c}{\partial \eta}&=&0,\qquad \text{on }\widetilde{F_{\pm}},\\
c&=&c_s,\qquad\text{on }\widetilde{S}.
\end{eqnarray}
The far-field boundary conditions ($t>0$) are:
\begin{equation}
\lim_{\stackrel{\xi\rightarrow\infty}{\eta=0}}c=c_\infty,
\end{equation}%
and
\begin{equation}
\lim_{z\rightarrow -\mathe^{\imag\alpha} }c=c_\infty.
\end{equation}%
Finally, the initial condition reads:
\begin{equation}
c=c_\infty,\qquad t=0.
\end{equation}
\label{eq:summaryMapped}%
\end{subequations}%
The diffusion equation is not conformally invariant as the mapped problem in Equation~\eqref{eq:diffusionMapped} contains a metric factor $|f'(z)|^2$.  Therefore, standard conformal mapping techniques for Laplace's equation do not apply.  However, we may divide the domain into two regions:
\begin{enumerate}[noitemsep]
\item A far field (or outer region) $r\gg a$ where the complex droplet geometry \textit{is not} important and the droplet may be treated like a point source.  In the outer region, standard techniques may be employed to solve the radially-symmetric \textit{time-dependent} diffusion equation in the physical space.
\item A near field (or inner region) where the complex geometry \textit{is} important but the droplet may be assumed to be locally in equilibrium.  In the inner region, the diffusion equation  reduces to \textit{time-independent} Laplace equation, which is conformally invariant and for which  standard methods from complex analysis may  be used to compute the vapour concentration.
\end{enumerate}
Then, the method of matched asymptotic expansions may be used to match the various solution forms, across an overlap region.

\subsection{Expression for the normal velocity of the free surface}

As the droplet is evaporating, the free surface moves.  Consequently, an expression for the normal velocity of the free surface $S$ must be found.
 Care must be taken here, as we wish to work in a Galilean frame in which the substrate faces $F_-$, $F$, and $F_+$ are stationary (lab frame).  
Relative to the lab frame, the origin of the $xy$ coordinate system moves with velocity
\begin{equation}
\bm{U}_c=(0,-\dot d),
\end{equation}
where the overdot denotes the derivative with respect to time.
Therefore, relative to the lab frame, we have:
\begin{equation}
V_n=\bm{U}_c\cdot\hatn+\dot R=-\dot d \sin\theta+\dot R.
\label{eq:Vn}
\end{equation}

To obtain an ordinary differential equation for the droplet size,  we use the expression $M=\rho_\ell a^2\mathcal{A}(\alpha)$ for the droplet mass.  Here, $a^2\mathcal{A}(\alpha)$ is the droplet area, where
\begin{equation}
\mathcal{A}(\alpha)=\frac{2\alpha-\sin(2\alpha)}{2\sin^2\alpha},\qquad \alpha \neq 0,\pi.
\end{equation}
Then, by conservation of mass,
\begin{equation}
\frac{\mathd}{\mathd t}\left[\rho_\ell a^2\mathcal{A}(\alpha)\right]=\underbrace{\int_{\theta_-}^{\theta_+}\myJ R\,\mathd \theta}_{=F_{2D}}.
\label{eq:dMdtDef}
\end{equation}
The quantity $F_{2D}$ is the integral flux in two dimensions, and depends on $t$, $\alpha$, and $a$.
To make further progress, we must specify the mode of evaporation as either constant radius (CR: $a=\text{Const.}$) or constant contact angle (CA: $\alpha=\text{Const.}$).  Here, we look at both modes of evaporation separately.  The following expressions rely on straightforward geometric reasoning and are derived in detail in the Supplementary Material.

\paragraph{CR Mode:}  In  the CA mode, $a=\text{Const}$, from which it can be shown that Equation~\eqref{eq:Vn} reduces to:
\begin{equation}
V_n(\theta,t)=-\frac{a\dot\alpha}{\sin^2\alpha} \left(\cos\alpha-\sin\theta\right).
\label{eq:VnCR2}
\end{equation}
Note that $V_n(\pi/2\pm \alpha,t)=0$, which is consistent with the CR assumption.  
From Equation~\eqref{eq:dMdtDef} we also have:
\begin{equation}
\rho_\ell a^2 \frac{\mathd\mathcal{A}}{\mathd \alpha}\dot\alpha=F_{2D}(t,\alpha,a),
\label{eq:massConsIntegral1}
\end{equation}
which is a closed-form ODE for $\alpha$.  The expression for $\dot\alpha$ may then be used to close the expression~\eqref{eq:VnCR2} for $V_n$.

\paragraph{CCA Mode:}  In the CCA mode,  $\tan\alpha=a/d=\text{Const}$.  Hence, Equation~\eqref{eq:Vn} can be re-written as:
\begin{equation}
V_n(\theta,t)=-\frac{\dot a}{\sin\alpha} \left(\cos\alpha\sin\theta-1\right).
\label{eq:VnCCA2}
\end{equation}
From Equation~\eqref{eq:dMdtDef} we also have:
\begin{equation}
2\rho_\ell [\mathcal{A}(\alpha)]a\dot a=F_{2D}(t,\alpha,a),
\label{eq:massConsIntegral2}
\end{equation}
which is a closed-form ODE for $a$.   The expression for $\dot a$ may then be used to close the expression~\eqref{eq:VnCCA2} for $V_n$.

\section{Vapour concentration in the atmosphere}
\label{sec:atmosphere}

As noted by Masoud and Felske~\cite{masoud2009analytical} and Yariv~\cite{yariv2023lifetime}, the diffusion equation for the vapour concentration has no steady-state solution in 2D and consequently, the full time-dependent diffusion equation must be considered.  In this section we outline how matched asymptotic expansions may be used to construct an approximate solution and hence, to find an expression for the evaporative flux.  Yariv's work is valid  in the slender-droplet approximation with $\alpha=0$; we extend this previous work such that our final expression for the evaporative flux is valid for all contact angles. 

\subsection{Outer solution}

In the far field, for $r\gg a$, we may work in the physical ($xy$) space since the precise droplet geometry is not important.   For the same reason, radial symmetry is assumed.  Consequently, the diffusion equation~\eqref{eq:model_all} reduces to:
\begin{equation}
\frac{\partial c}{\partial t}=D_v\frac{1}{r}\frac{\partial }{\partial r}\left(r\frac{\partial c}{\partial r}\right).
\label{eq:diffusionRad}
\end{equation}
We introduce the Laplace transform of $c$:
\begin{equation}
\hatcL(r,s)=\int_0^\infty c(r,t)\mathe^{-st}\mathd t.
\end{equation}
The diffusion equation~\eqref{eq:diffusionRad} may itself be Laplace-transformed to give:
\begin{equation}
c_\infty+s\hatcL=D_v\frac{1}{r}\frac{\partial}{\partial r}\left(r\frac{\partial\hatcL}{\partial r}\right).
\label{eq:diffusionL}
\end{equation}
The solution to Equation~\eqref{eq:diffusionL} is:
\begin{equation}
\hatcL(r,s)=\frac{c_\infty}{s}+A(s)K_0\left( \frac{s^{1/2}r}{D_v^{1/2}}\right),
\end{equation}
where $K_0$ is the modified Bessel function of the second kind.  We use the small-argument approximation of $K_0$ to obtain:
\begin{equation}
\hatcL(r,s)\sim \frac{c_\infty}{s} +A(s)\left[\ln \left(\frac{2D_v^{1/2}}{rs^{1/2}}\right)-\gamma_E\right],
\label{eq:hatcLSln}
\end{equation}
valid as $r\rightarrow 0$.  Here, $\gamma_E$ is the Euler--Mascheroni constant.  In the mapped space, Equation~\eqref{eq:hatcLSln} can be written as:
\begin{equation}
\widehat{c}_{outer}(s,r)= \frac{c_\infty}{s} +A(s)\left[ \ln|\zeta-\zeta_0|+\ln\left(\frac{2 (D_v/s)^{1/2}}{|z_+-z_-|}\right)\right],
\label{eq:cOuterLaplaceMatch}
\end{equation}
where $\zeta_0=-\mathe^{\imag\alpha}$.  In equation~\eqref{eq:cOuterLaplaceMatch}, we have added the  subscript `outer'  to $\widehat{c}$, emphasizing that this particular form will be used to match the outer solution for $c$ with the inner solution.

\subsection{Inner solution}

\begin{figure}
\centering
\begin{tikzpicture}[scale=1.2]

    \def\L{5.0}
    \def\myalpha{120}

		\fill[
    pattern=north east lines,
    pattern color=blue!50
    ]
    (0,0)
    -- ({\L*cos(\myalpha)}, {\L*sin(\myalpha)})
    arc[start angle={\myalpha}, end angle=0, radius=\L]
    -- cycle;

    \draw[->] (-1.0,0) -- (1.2*\L,0);
    \draw[->] (0,-0.5) -- (0,5.5);
		

    \draw[very thick,blue,->] (0,0) -- ({1.1*\L*cos(\myalpha)}, {1.1*\L*sin(\myalpha)});
		\node [draw,above,fill=white] at (2.4,1.6) {$\widetilde{\Omega_a}$}; 
		\node [draw,above] at (-1.55,1.8) {$\widetilde{S}$}; 
		\node [above] at (-1.2,1) {$c=c_s$};

    \draw[very thick,green]
        (0,0) -- (1.5,0);
		\node [draw,below] at (0.75,-0.1) {$\widetilde{F_+}$}; 

	  \draw[very thick,black,->]
		   (1.5,0)--(1.1*\L,0);
		\node [draw,below] at (0.5*\L,-0.1) {$\widetilde{F_-}$}; 
		
		\fill[red] (1.5,0) circle (0.1cm);
		\fill[red] (0,0) circle (0.1cm);
		
		\node [draw,below,red,align=center] at (1.5,-0.8) {Logarithmic \\ singularity}; 
		\draw[red,->] (1.5,-0.8) -- (1.5,-0.1);
		
		\node[above,
    align=center,
    fill=white,
    inner sep=2pt] at (1.5,0.2) 
		{{\textcolor{red}{(1,0)}}};
		
		\node [below] at (4,-0.15) {$\frac{\partial c}{\partial y}=0$};
		
    \node [below] at (0.1,-0.8) {$\frac{\partial c}{\partial y}=0$};
		\draw[->] (0.1,-0.8) -- (0.3,-0.1);
		
		\draw
    (2.2,0)
    arc[start angle=0, end angle={120}, radius=2.2]
    node[midway,  above] {$\beta$};

\end{tikzpicture}
\caption{Modified mapped domain, using the polar coordinate $\varphi$.  Wedge angle: $\beta=\pi-\alpha$}
\label{fig:mappedDomainNew}
\end{figure}
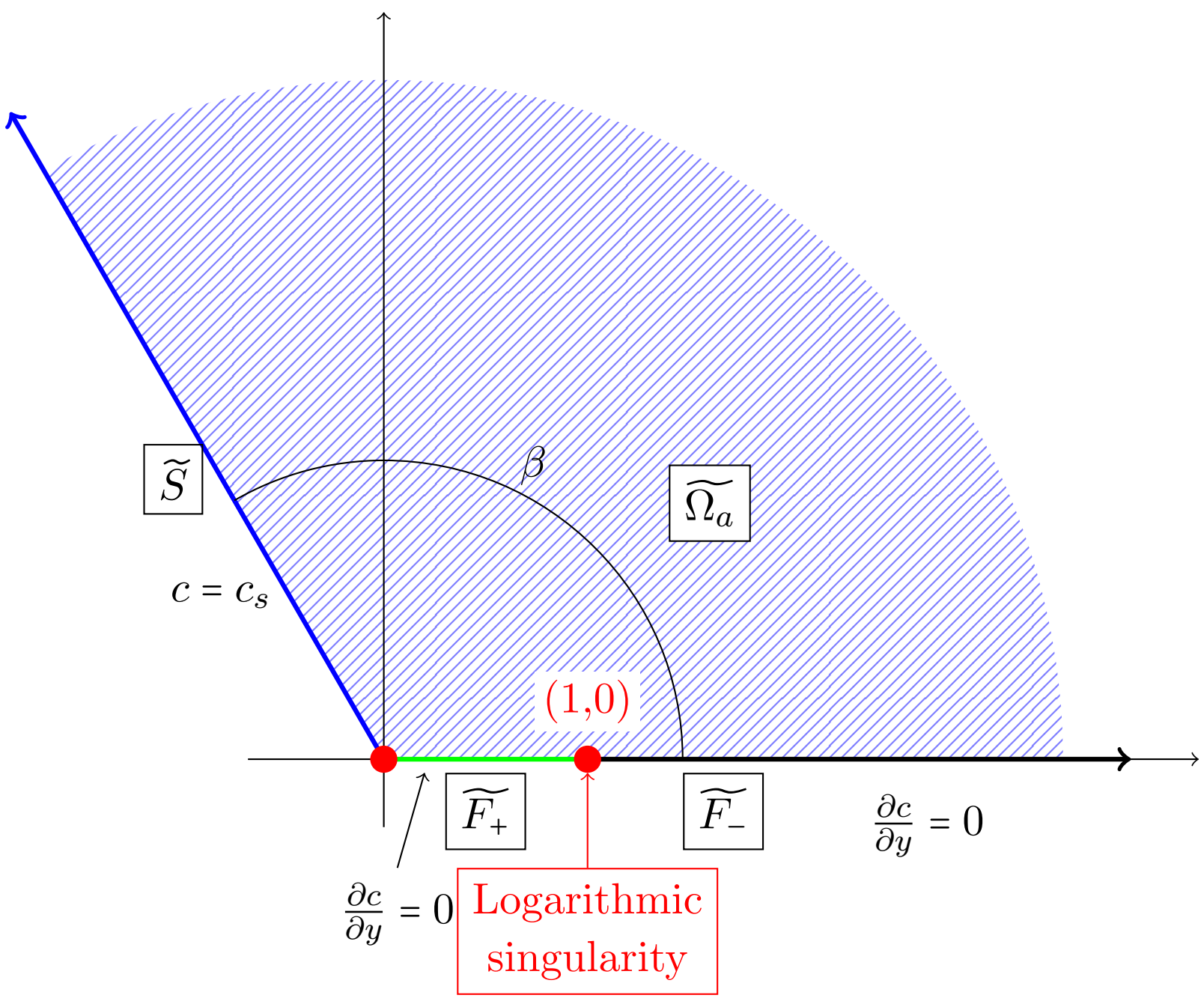

As noted in Section~\ref{sec:theory}, in the inner region $r\apprge R$, the vapour concentration may be assumed to be in steady state.  Consequently, the diffusion equation~\eqref{eq:model_all} reduces to $\nabla_{\xi\eta}^2c=0$.  To simplify computations, we  use the variable $\varphi=\arg(\zeta)-(\alpha+\pi)$, corresponding to the wedge geometry in Figure~\ref{fig:mappedDomainNew}.  In terms of the $(\rho,\varphi)$ polar coordinates introduced in Section~\ref{sec:theory}, this is:
\begin{equation}
\varphi=\phi-(\alpha+\pi).
\end{equation}
We also introduce the auxiliary angle $\beta$, where
\begin{equation}
\beta=\pi-\alpha.
\end{equation}
Thus, the wedge angle in Figure~\ref{fig:mappedDomainNew} is at $\varphi=\beta$.  To absorb the inhomogeneous boundary condition along the ray $\varphi=\beta$, we use the trial solution $c=c_s+v$.  The equation to be solved now reads:
\begin{subequations}
\begin{eqnarray}
\nabla^2 v&=&0,\\
v&=&0,\qquad \varphi=\beta,\\
\frac{\partial v}{\partial \varphi}&=&0,\qquad \varphi=0.
\end{eqnarray}
Laplace's equation in an infinite wedge does not have a unique solution, as there are not enough boundary conditions to constrain all of the constants of integration~\cite{strauss2007partial}.  To obtain a unique solution, we apply the extra condition:
\begin{equation}
v\sim - C\ln |\rho-1|,\qquad \text{as }\rho\rightarrow 1,
\label{eq:lnsln}
\end{equation}
where $C$ is a free (positive) constant.
Condition~\eqref{eq:lnsln} ensures that the complete inner solution matches up with the outer solution in the overlap region.
\label{eq:laplaceWedge}
\end{subequations}

We solve Equation~\eqref{eq:laplaceWedge} using separation of variables, $v=F(\rho)\Phi(\varphi)$.  
After lengthy but standard steps as outlined in the Supplementary Material, we obtain:
\begin{equation}
c_{inner}(\rho,\varphi,t)=c_s-\frac{Q(t)}{ D_v(\pi-\alpha)}\sum_{n=0}^\infty \frac{1}{\lambda_n}\cos(\lambda_n\varphi)F_n(\rho),
\label{eq:cInnerFinal}
\end{equation}
where
\begin{equation}
F_n(\rho)=\begin{cases}\rho^{\lambda_n},&0<\rho\leq 1,\\
                       \rho^{-\lambda_n},&\rho\geq 1,\end{cases} 
\end{equation}
and
\begin{equation}
\lambda_n=\frac{\left(n+\tfrac{1}{2}\right)\pi}{\beta},\qquad n\in \{0,1,2,\cdots\}.
\end{equation}
In Equation~\eqref{eq:cInnerFinal} we have  used the subscript `inner' to indicate that the expression for $c$ so derived is the inner solution, valid near the droplet free surface.  In the same manner, $Q(t)$ is a free term (possibly time-dependent) which must be introduced as a prefactor to allow the inner and outer solutions to be matched across an overlap region.  To facilitate the matching, we write $c_{inner}$ in terms of its Laplace transform:
\begin{equation}
\widehat{c}_{inner}(\rho,\varphi,s)=\frac{c_s}{s}-\frac{\widehat{Q}(s)}{ D_v\beta}\sum_{n=0}^\infty \frac{1}{\lambda_n}\cos(\lambda_n\varphi)F_n(\rho).
\label{eq:cInnerFinalL}
\end{equation}
where $\widehat{Q}(s)$ is the Laplace transform of $Q(t)$.

\subsection{Matching inner and outer solutions}

Equation~\eqref{eq:cInnerFinalL} may be evaluated on the boundary faces $F_{\pm}$ in the limit of large $|x|$ to give:
\begin{equation}
\widehat{c}_{inner}(r,s)\sim \frac{c_s}{s}+\frac{\widehat{Q}(s)}{\pi D_v} \left[ \ln|1-\rho|-\ln(4\beta/\pi)\right].
\label{eq:cInnerFinalLAsy}
\end{equation}
The passage from the sum in Equation~\eqref{eq:cInnerFinalL} to the logarithmic expression in Equation~\eqref{eq:cInnerFinalLAsy} is worked out in detail in the Supplementary Material.
Referring back to Equation~\eqref{eq:cOuterLaplaceMatch}, on the boundary faces $F_{\pm}$ and in the limit of small $|x|$, we have:
\begin{equation}
\widehat{c}_{outer}= \frac{c_\infty}{s} +A(s)\left[ \ln|\zeta-\zeta_0| +\ln \left(\frac{2(D_v /s)^{1/2}}{|z_+-z_-|}\right)-\gamma_E+\cdots\right],\qquad \text{as }\zeta\rightarrow \zeta_0.
\label{eq:cOuterLAsy}
\end{equation}
Since $\ln|\zeta-\zeta_0|=\ln|1-\rho|$ we  have, matching other terms in the inner and outer solution in the overlap region,
\begin{equation}
\frac{c_s}{s}+\frac{\widehat{Q}(s)}{\pi D_v}\left[\cancel{\ln|1-\rho|}-\ln(4\beta/\pi)\right]=\frac{c_\infty}{s}+A(s)\left[\cancel{\ln|\zeta-\zeta_0|}+\ln \left(\frac{2(D_v /s)^{1/2}}{|z_+-z_-|}\right)-\gamma_E\right].
\end{equation}
where the cancellation occurs provided we take:
\begin{equation}
A(s)=\frac{\widehat{Q}(s)}{\pi D_v}.
\end{equation}
This gives a closed expression for $\widehat{Q}(s)$:
\begin{equation}
\widehat{Q}(s)=\frac{1}{s}\frac{(c_s-c_\infty)\pi D_v}{\ln  \left(\frac{2(D_v /s)^{1/2}}{|z_+-z_-|}\right)+\ln(4\beta/\pi)-\gamma_E}.
\end{equation}
Alternatively, we may write this as:
\begin{equation}
\widehat{Q}(s)=\frac{1}{s}\frac{(c_s-c_\infty)\pi D_v}{\tfrac{1}{2}\ln  \left(\frac{16(D_v /s)}{a^2}\right)+\ln(\beta/\pi)-\gamma_E}.
\end{equation}
This result (with non-zero $\alpha$) is the same in form as Yariv's result for $\alpha=0$~\cite{yariv2023lifetime}, provided we take:
\begin{equation}
\gamma_{eff}=\gamma_E-\ln(\beta/\pi).
\end{equation}
Consequently, we may jump to an expression for $Q(t)$ in real space:
\begin{equation}
Q(t)=\tfrac{1}{2}(c_s-c_\infty)\pi D_v\int_0^\infty \mathe^{-\lambda t} \frac{1}{\left[\tfrac{1}{2}\ln  \left(\frac{16(D_v /\lambda)}{a^2}\right)-\gamma_{eff}\right]^2+\tfrac{1}{4}\pi^2}\frac{\mathd \lambda}{\lambda}.
\label{eq:yarinF}
\end{equation}
The singularity in the integrand of Equation~\eqref{eq:yarinF} is cosmetic and may be removed by appropriate $u$-substitution.  Furthermore, the integral may be evaluated using Laplace's asymptotic method for integrals  in the limit of large $t$ (see Supplementary Material) to give:
\begin{equation}
Q(t)\sim  \frac{\pi D_v(c_s-c_\infty)}{\ln \left(\frac{\sqrt{D_v t}}{a}\right)+\ln(4\beta/\pi)-\tfrac{1}{2}\gamma_E},\qquad \text{as }t\rightarrow\infty.
\label{eq:QtAsy}
\end{equation}

\subsection{Expression for the evaporative flux}

From Equation~\eqref{eq:Jdef}, we may work out the evaporative flux for a 2D cylindrical droplet, using the inner solution to compute the vapour-concentration gradient.  Taking account of the metric factor induced by transforming between $xy$ coordinates and $\xi\eta$ coordinates, the evaporative flux is:
\begin{equation}
\myJ=-D_v(\hatn\cdot\nabla_{xy}c)_{r=R}=D_v\left|\frac{\mathd f}{\mathd z}\right|_{\eta=0}\left(\frac{\partial c}{\partial \eta}\right)_{\eta=0}.
\label{eq:J2Dexpr1}
\end{equation}
Using the property of the metric factor $|\mathd f/\mathd z|$ in Equation~\eqref{eq:dfdzMetric} this becomes:
\begin{equation}
\myJ=D_v\frac{\rho^2+2\rho\cos\alpha+1}{2a}\left(\frac{\partial c}{\partial \eta}\right)_{\eta=0}.
\label{eq:J2Dexpr2}
\end{equation}
The inner solution~\eqref{eq:cInnerFinal} is substituted into~\eqref{eq:J2Dexpr2} and evaluated at $\eta=0$ ($\varphi=\beta$) to produce a final closed-form expression for the evaporative flux, as a function of the polar angle $\theta$ in the physical space:
\begin{subequations}
\begin{equation}
\myJ(\theta)=\frac{Q(t)}{2a\beta}\left(\rho^2+2\rho\cos\alpha+1\right)\frac{1}{\rho}\frac{\rho^{\pi/2\beta}}{1+\rho^{\pi/\beta}},
\end{equation}
where $\rho$ is constrained:
\begin{equation}
\rho=\mathcal{R}(\theta),\qquad \mathcal{R}(\theta)=-\frac{\sin\left(\frac{\theta-\theta_+}{2}\right)}{\sin\left(\frac{\theta-\theta_-}{2}\right)},\qquad \theta_+<\theta<\theta_-.
\end{equation}%
\label{eq:Jfinal}%
\end{subequations}%

\subsection{Analysis of the evaporative flux}

We look at the limiting behaviour of the evaporative flux~\eqref{eq:Jfinal} near the triple point $\theta\rightarrow \theta_+$.  Referring back to Equation~\eqref{eq:Jfinal}, we have:
\begin{equation}
\myJ\sim \frac{Q(t)}{2a\beta}\rho^{\frac{\pi}{2\beta}-1},\qquad \text{as }\theta\rightarrow \theta_+.
\end{equation}
Using a Taylor approximation to $\rho=\mathcal{R}(\theta)$ valid as $\theta\rightarrow \theta_+$, we obtain:
\begin{equation}
\rho\sim \frac{\theta-\theta_+}{2\sin\alpha},\qquad \text{as }\theta\rightarrow \theta_+.
\end{equation}
Let $\ell$ denote the  arc length from the triple point $z_+$ along the free surface.  Therefore, $\ell=R(\theta-\theta_+)$ and
\begin{equation}
\frac{\theta-\theta_+}{2\sin\alpha}=\frac{R(\theta-\theta_+)}{2R\sin\alpha}=\frac{\ell}{2a},
\end{equation}
hence
\begin{equation}
\rho \sim \frac{\ell}{2a}\qquad\text{as }z\rightarrow z_+.
\label{eq:rhoLim}
\end{equation}
Therefore, in the limit as $\ell\rightarrow 0$, we have:
\begin{equation}
\myJ \sim \frac{Q(t)}{2a(\pi-\alpha)}\left(\frac{\ell}{2a}\right)^{\overbrace{\frac{\pi}{2(\pi-\alpha)}}^{=\lambda_{2D}}-1}.
\label{eq:Jlim2D}
\end{equation}
This can be compared with the expression for the flux in the three-dimensional (spherical cap) problem~\cite{wilson2023evaporation}, which we recall here as:
\begin{equation}
J_{3D}\sim (a-r)^{\lambda_{3D} -1},\qquad \text{as }r\uparrow a,
\label{eq:J3D}
\end{equation}
where $\lambda_{3D}=\pi/[2(\pi-\alpha)]$.  We may therefore unambiguously write:
\begin{equation}
\lambda_{2D}=\lambda_{3D}=\frac{\pi}{2(\pi-\alpha)}:=\lambda,
\end{equation}
a result first obtained by Yarin \textit{et al.}~\cite{yarin2006lines}.   Out work builds up on this prior work because it is independent of the (arbitrary) far-field location $R$ and also, in view of the relativity simple closed-from expression for the evaporative flux in~\eqref{eq:Jfinal}.  To enable further progress with respect to the prior work, we next introduce a simple closed-form expression for the integral flux, which is useful in formulating the dynamics of the evaporating droplet in both the CR and CCA modes.

\subsection{Integral flux}

We derive a closed-form expression for the integral flux $F_{2D}$ previously introduced in Equation~\eqref{eq:dMdtDef}.  In this expression, the flux $\myJ$ is integrated along a line element $\mathd s_{xy}=R\mathd\theta$.
By conformal mapping, this is related to the line element in the $\xi\eta$ space:
\begin{equation}
\mathd s_{xy}=\frac{\mathd s_{\xi\eta}}{|\mathd f/\mathd z|}.
\end{equation}
Therefore:
\begin{eqnarray*}
F_{2D}&=&\int_{\theta_-}^{\theta_{+}}\myJ \,R\mathd \theta,\\
 &\stackrel{\text{Eq.~\eqref{eq:Jfinal}}}{=}&\frac{Q(t)}{2a\beta}\int_{\theta_-}^{\theta_{+}}\frac{\rho^2+2\rho\cos\alpha+1}{\rho}\frac{\rho^{\frac{\pi}{2\beta}}}{1+\rho^{\frac{\pi}{\beta}}}\mathd s_{xy},\\
 &\stackrel{\text{Eq.~\eqref{eq:dfdzMetric}}}{=}& \frac{Q(t)}{2a\beta}\int_{\theta_-}^{\theta_{+}}\frac{2a\cancel{|\mathd f/\mathd z|}}{\rho}\frac{\rho^{\frac{\pi}{2\beta}}}{1+\rho^{\frac{\pi}{\beta}}}\frac{\mathd s_{\xi\eta}}{\cancel{|\mathd f/\mathd z|}}.
\end{eqnarray*}
We recall that $\zeta=\xi+0\imag$ on the mapped free surface $\widetilde{S}$, and that $\theta\in (\theta_-,\theta_+)$ corresponds to $\xi\in(0,\infty)$.  This gives:
\begin{equation}
F_{2D}=\frac{Q(t)}{\beta}\underbrace{\int_0^\infty \frac{\xi^{\mu-1}}{1+\xi^{2\mu}}\mathd \xi}_{=\frac{\pi}{2\mu}}.
\end{equation}
Using the substitution $u=\xi^\mu$ gives the stated result, hence:
\begin{equation}
F_{2D}=Q(t).
\label{eq:F2Dfinal}
\end{equation}
Hence, all contact-angle information is integrated out; the only information on contact angle appears through the term $Q(t)$.  Therefore, from the integral~\eqref{eq:yarinF} and Equation~\eqref{eq:QtAsy}, we have:
\begin{multline}
F_{2D}=Q(t)\sim  \frac{\pi D_v(c_s-c_\infty)}{L(t)}\,\text{where } L(t)=\ln \left(\frac{\sqrt{D_v t}}{a}\right)+\ln(4\beta/\pi)-\tfrac{1}{2}\gamma_E,\\
 \text{as }\ln\left(D_v t/a^2\right) \rightarrow\infty.
\label{eq:myF}
\end{multline}
Equation~\eqref{eq:myF} is asymptotically valid as $\ln ( D_v t/a^2)\rightarrow \infty$.  This is in contrast with traditional asymptotic expansions, which use $t$ itself as the asymptotic parameter.  Since $\ln (t/t_0)\ll t/t_0$ for large $t$ and fixed $t_0$ (such that $t/t_0$ is dimensionless), an improved asymptotic expression requiring the next-order correction in the integral~\eqref{eq:yarinF} is required, giving:
\begin{equation}
F_{2D}=Q(t)\sim  \frac{\pi D_v(c_s-c_\infty)}{L(t)+\tfrac{\pi^2}{24}\left(\frac{1}{L(t)-\tfrac{1}{2}\gamma_E}\right)},\qquad \text{as }\ln\left(D_v t/a^2\right) \rightarrow\infty.
\label{eq:myF2}
\end{equation}
The asymptotic expressions are only valid at late times.  Indeed, the expression~\eqref{eq:myF2} has a singularity at an early time; however, the equation has no physical meaning at such early times.

The similarity between 2D and 3D holds up locally, near the vicinity of the triple point, where the expression for the evaporative flux $J$ has the same form in 2D and 3D.  However, the similarity ends when the global picture is considered, that is, when the evaporative flux is integrated over the circular segment in 2D or the spherical cap in 3D.  The resulting integral expressions have very distinctive behaviours.  In the 2D case, the integral flux  is given by Equations~\eqref{eq:F2Dfinal}--\eqref{eq:myF2} and depends only very weakly on contact angle, through a logarithmic term.  In contrast, the 3D integral flux contains a strong contact-angle dependence, whose exact form is given in Reference~\cite{wilson2023evaporation} (Equation~(10) therein).

\section{Numerical Computation of the Evaporative Flux}
\label{sec:numerics}

In this section, we carry out numerical simulations to  validate the expressions for the integral flux $F_{2D}$ (specifically, Equations~\eqref{eq:yarinF}, and~\eqref{eq:F2Dfinal}--\eqref{eq:myF2}).  Expression~\eqref{eq:yarinF} is derived under the assumption that the vapour domain can be partitioned into inner and outer regions and the diffusion equation can be solved using the method of matched asymptotic expansions; the expressions~\eqref{eq:F2Dfinal}--\eqref{eq:myF2} require the extra assumption that the late-time limit $t\gg a^2/D_v$ has been reached.  Therefore, in this section we use numerical simulations to test these assumptions.

We use the OpenFOAM C++ toolbox for solving the time-dependent diffusion equation.   The toolbox allows the user to develop customized numerical solvers for continuum mechanics problems, in particular computational fluid dynamics.  However, the toolbox comes equipped with a range of already-built solvers and tutorials, meaning that simple test-case simulations can be set up with reduced effort (i.e. compared to developing custom-built solvers). One such solver is \texttt{laplacianFOAM}, which simulates the time-dependent diffusion equation in a user-specified computational domain.  For this purpose, we have developed a custom computational mesh using the \texttt{snappyhexmesh} tool, which accounts for the curved geometry in Figure~\ref{fig:schematic1}.  
\begin{figure}[htb]
	\centering
		\includegraphics[width=0.8\textwidth]{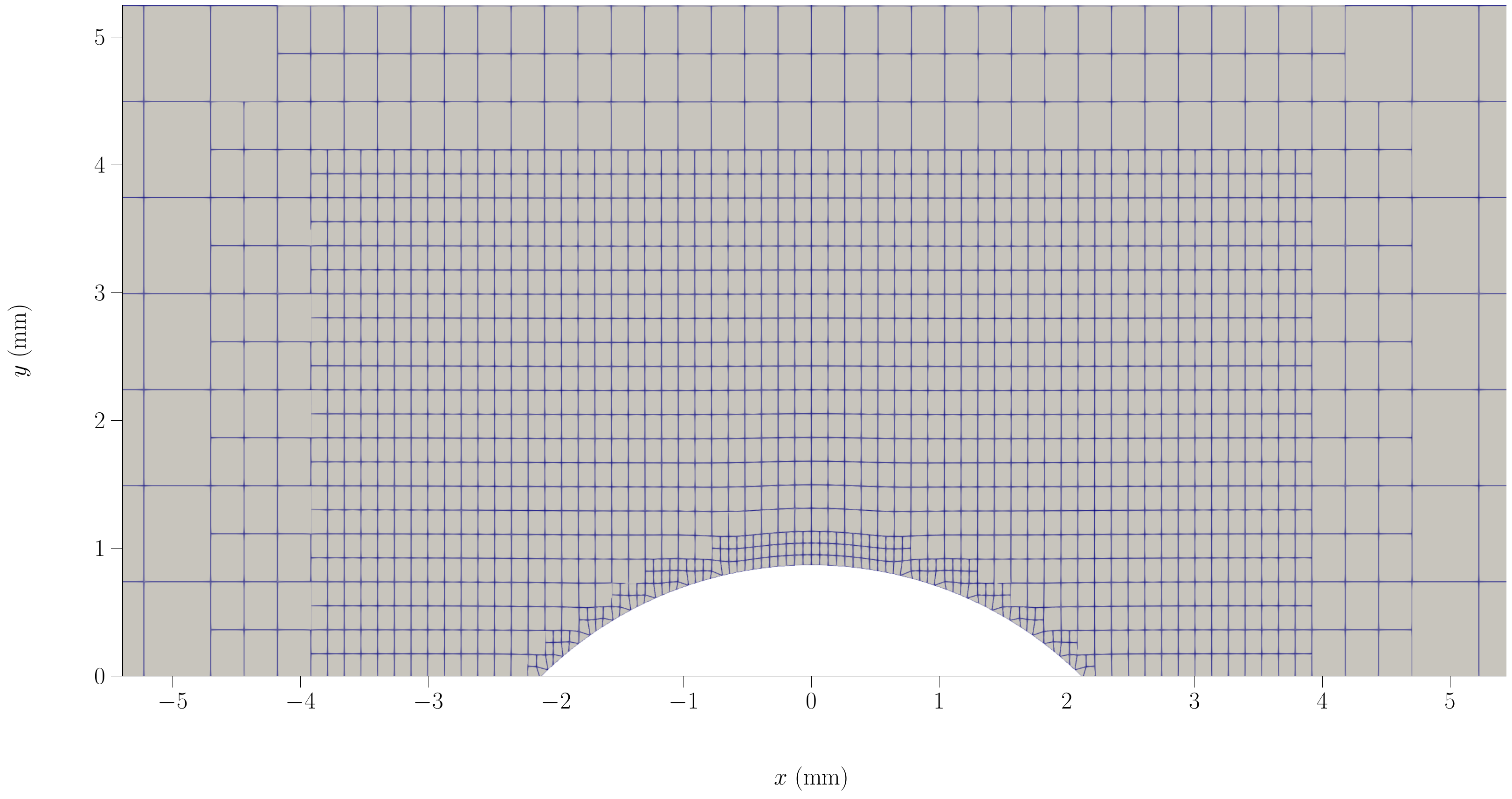}
		\caption{Schematic diagram showing the numerical mesh for the simulation}
	\label{fig:mesh-zoom}
\end{figure}

A box-shaped computational domain is used in the simulations, with sides of  length $L_x=0.12\,\mathrm{m}$ and $L_y=0.06\,\mathrm{m}$.   As OpenFOAM requires a three-dimensional domain, the domain involves a third dimension into the plane of the page; this is taken to have thickness $\Delta z=5.625\times 10^{-5}\,\mathrm{m}$, giving an effectively two-dimensional domain.   The \texttt{snappyhexmesh} OpenFOAM utility is then used effectively to `cut out' the cylindrical-cap shape, leaving the computational domain shown in Figure~\ref{fig:mesh-zoom}.  Both the computational mesh, and the full OpenFOAM case, are provided in an online repository which accomapnies this work~\cite{Quigley_HeatEvaporationFOAM}.

We  set $c_{s,3D} = 1\,\mathrm{kg}\cdot\mathrm{m}^{-3}$ and $c_{\infty,3D} = 0\,\mathrm{kg}\cdot\mathrm{m}^{-3}$ in the simulations.  The concentrations have the appropriate units for a three-dimensional simulation; the exact values are not important as concentrations will be divided by $(c_{s,3D}-c_{\infty,3D})\Delta z$ for comparison with 2D calculations.    We set $D_{v} = 2.42 \times 10^{-5}\,\mathrm{m}^2\mathrm{s}^{-1}$, corresponding to the diffusivity of water vapour in air at $20^\circ\mathrm{C}$.  For the other droplet parameters, we use a contact angle $\alpha=45^\circ$ and a circular radius $R=0.003\,\mathrm{m}$ giving a droplet footprint $a = 2.12 \times 10^{-3}\,\mathrm{m}$, following $a = R \sin\alpha$.  Local mesh refinement is applied in the vicinity of the droplet surface.  The number of computational cells in the mesh is $N = 19,796$.  A mesh-refinement study using a simulation with $N=78,558$ cells gives the same results.  Therefore, throughout this section, we present simulation results with  $N = 19,796$.

\begin{figure}[H]
  \centering
	\subfigure[$t = 0.1\,\mathrm{s}$]{\includegraphics[width=0.85\textwidth]{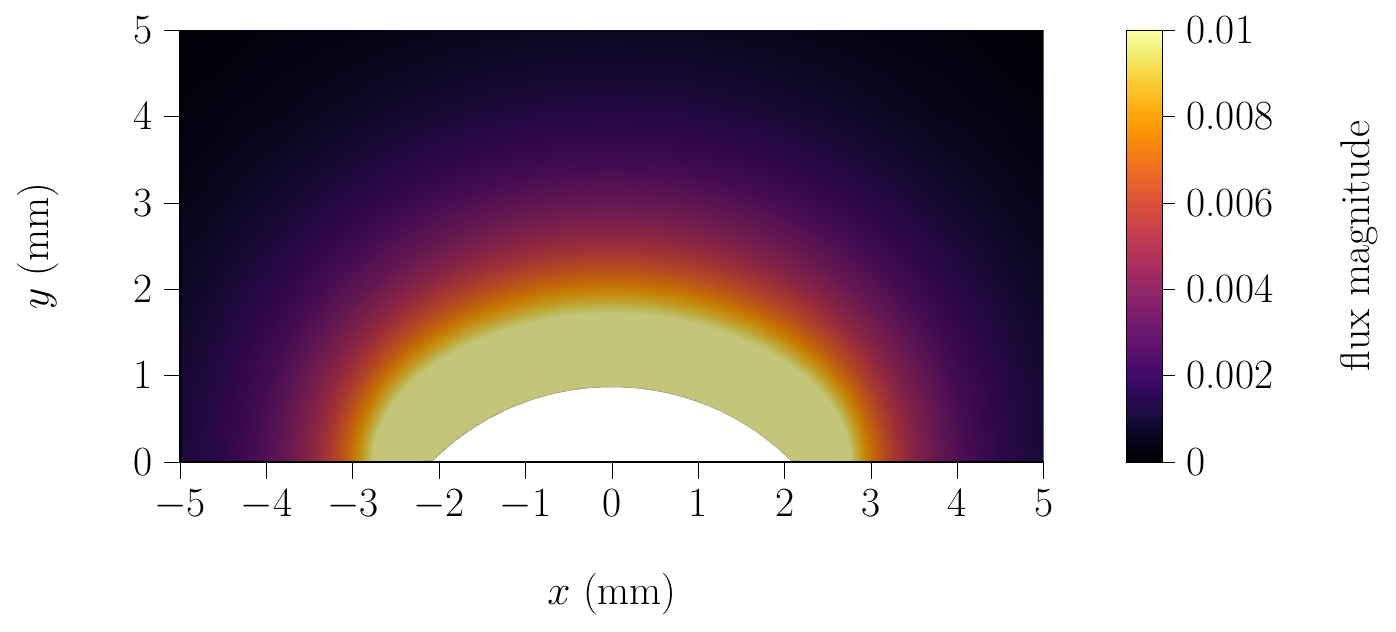}}
	\subfigure[$t = 2\,\mathrm{s}$]{\includegraphics[width=0.85\textwidth]{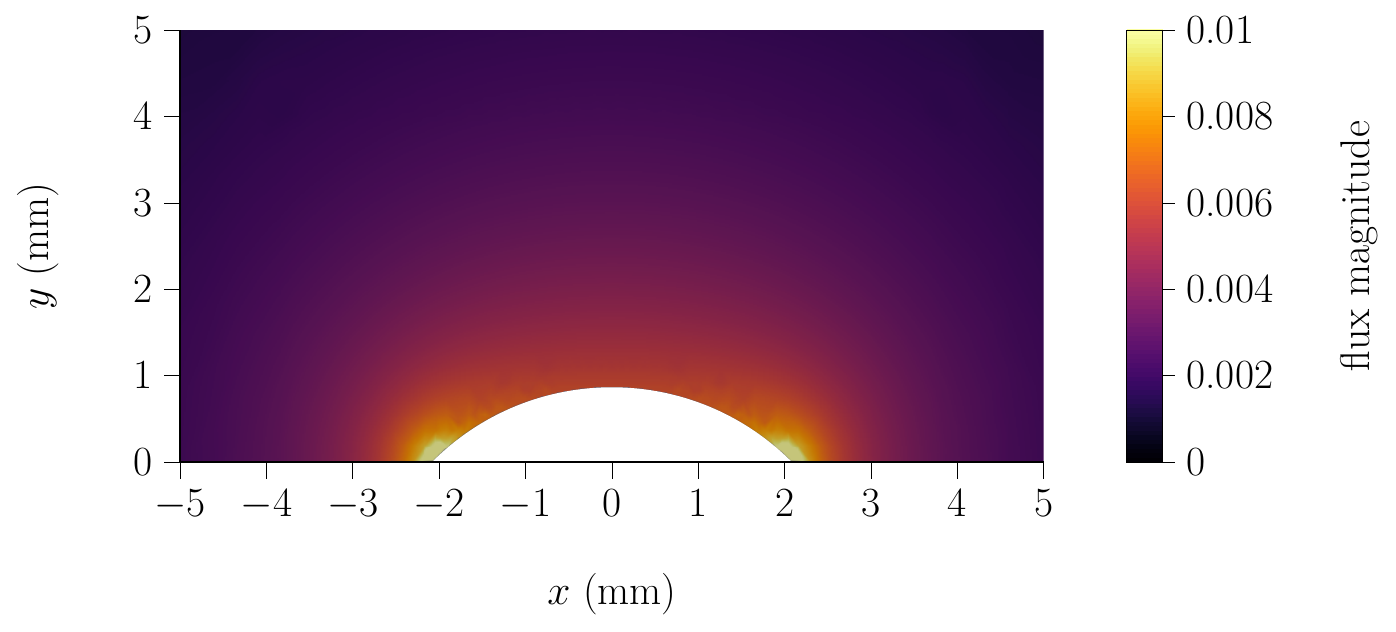}}
	\subfigure[$t = 30\,\mathrm{s}$]{\includegraphics[width=0.85\textwidth]{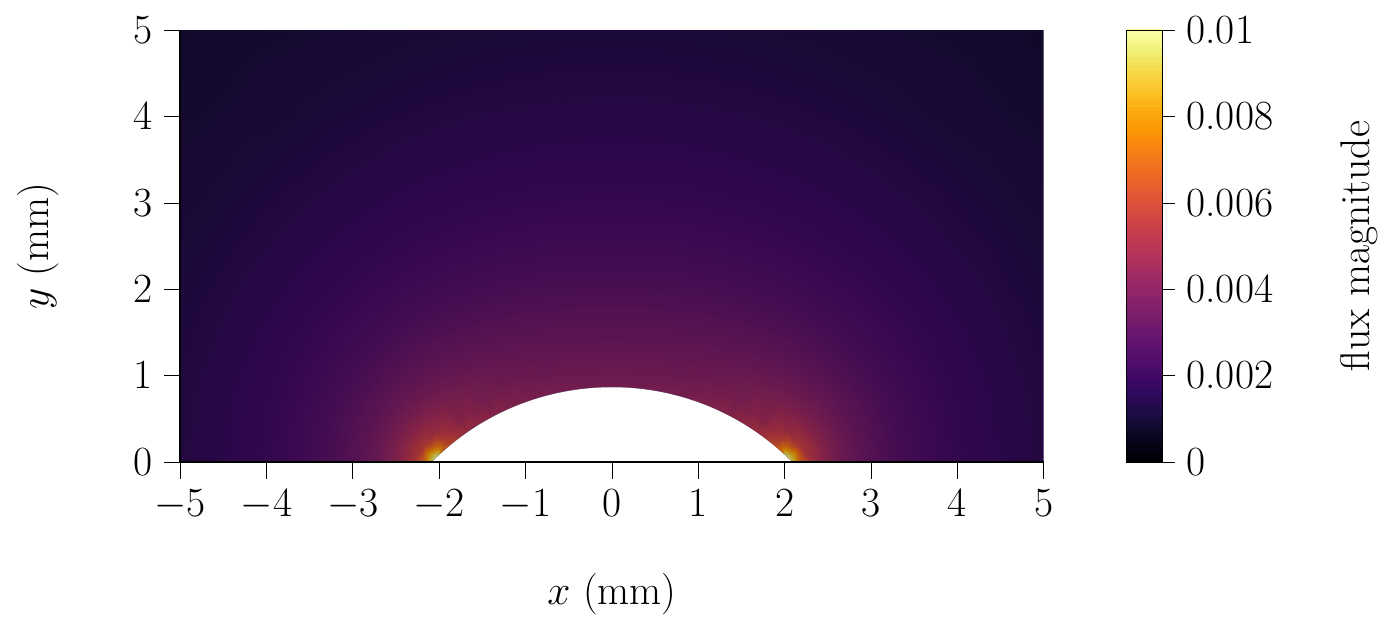}}
  \caption{Flux magnitude $D_v| \nabla c|$ at different times during the simulation.  As the simulations are carried out in a three-dimensional computational domain, the units of the flux magnitude  in the plots are $\mathrm{kg}\cdot\mathrm{m}^{-2}\cdot\mathrm{s}^{-1}$.}
  \label{fig:flux-grid}
\end{figure}

A first set of results is shown in Figure~\ref{fig:flux-grid}.  Shown in this figure are snapshots of the flux magnitude $D_v|\nabla c|$ over time.  The snapshots show a concentration of flux at the triple points, consistent with the analysis in Section~\ref{sec:atmosphere}.  However, the flux magnitude remains finite in the simulations, even at the triple points.  This is in contrast to the analytical calculations in  Section~\ref{sec:atmosphere}.  In that section, the flux at the triple points is obtained at late times, by solving Laplace's equation.  Laplace's equation with the chosen boundary conditions admits a flux singularity at the triple points.   However, the singularity is `integrated out' when the integral flux is considered.  Therefore, if exact quantitative agreement between the numerical simulations and the asymptotic expressions is to be found, it will be via  the integral flux derived in  Section~\ref{sec:atmosphere}.

\begin{figure}[htb]
	\centering
		\includegraphics[width=0.9\textwidth]{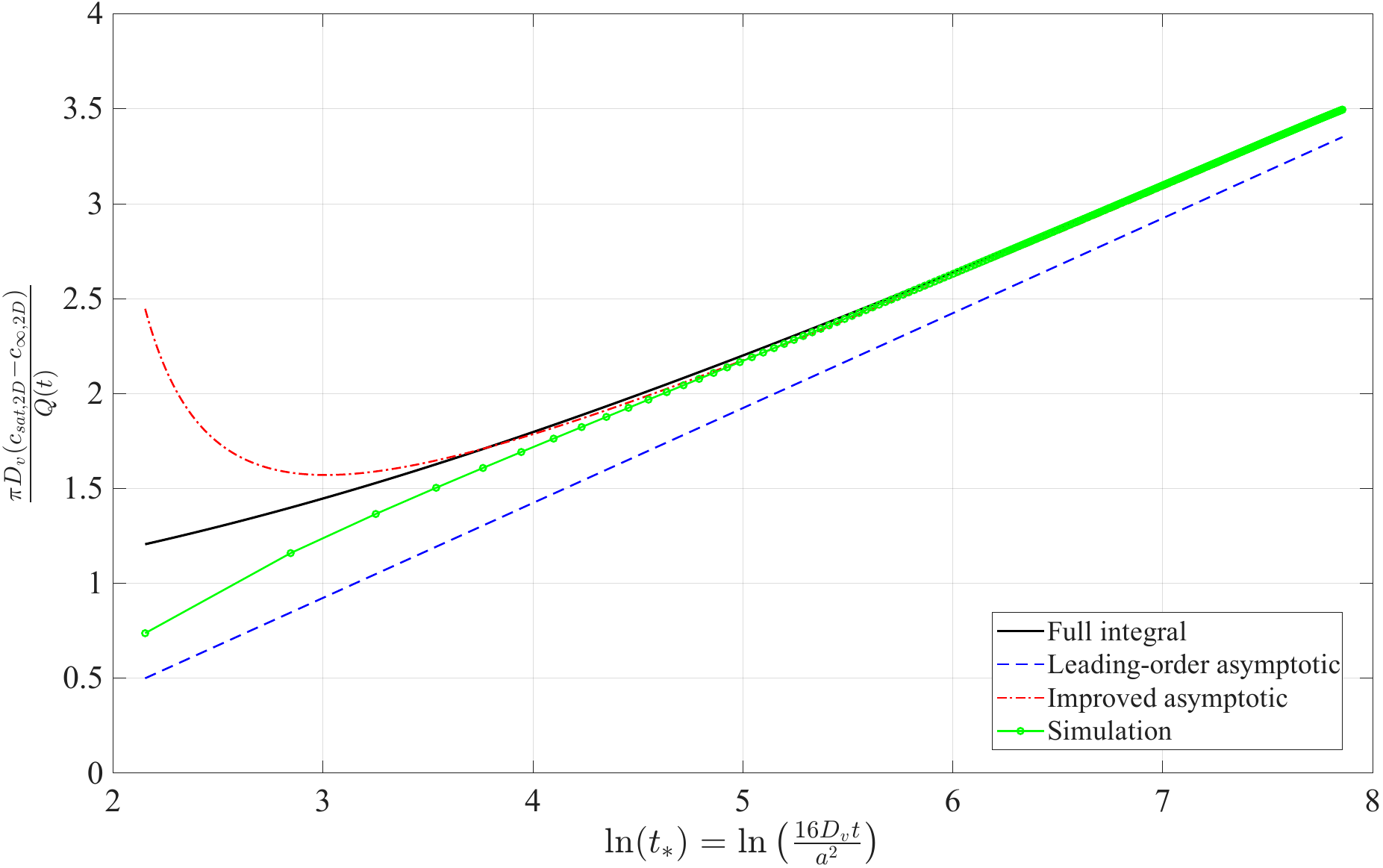}
		\caption{Comparison between the numerical value of the integral flux, and the various analytical expressions derived in Section~\ref{sec:atmosphere}.  Legend key: `Full Integral' refers to the expression based on the Yariv-type integral in Equation~\eqref{eq:yarinF}.  `Leading-order' asymptotic refers to the formula in Equation~\eqref{eq:myF}.  `Improved asymptotic' refers to the formula in in Equation~\eqref{eq:myF2}.  The asymptotic expressions are only valid at late times.  To ensure dimensional consistency, for the 3D simulations, we use $c_{s,2D}=c_{s,3D}\Delta z$ and $c_{\infty,2D}=c_{\infty,3D}$, where $\Delta z=5.625\times 10^{-5}\,\mathrm{m}$ is the thickness of the third dimension in the simulations.  Here, as in the main text,  $c_{s,3D} = 1\,\mathrm{kg}\cdot\mathrm{m}^{-3}$ and $c_{\infty,3D} = 0\,\mathrm{kg}\cdot\mathrm{m}^{-3}$. }
	\label{fig:sim-analytical-comparison}
\end{figure}

To test this hypothesis, and to validate the expressions for the integral flux $F_{2D}$ (specifically, Equations~\eqref{eq:yarinF}, and~\eqref{eq:F2Dfinal}--\eqref{eq:myF2}), we look at the integral flux from the numerical simulations.  The integral flux  can be computed directly in OpenFOAM using the \texttt{areaNormalIntegrate} function.   
%
%
%
%
%
Results are shown in Figure~\ref{fig:sim-analytical-comparison}.  There is exact agreement between the numerical value of the integral flux and the analytical expression  based on Yariv's  integral expression (Equation~\eqref{eq:yarinF}).  For sufficiently late time, the improved asymptotic expression~\eqref{eq:myF2} for the integral flux also agrees well with the simulation value.  In contrast, the basic asymptotic expression~\eqref{eq:myF} is unable to capture the behaviour of the integral flux over the $30\,\mathrm{s}$ time interval covered by the simulation.

\newpage

\section{Flow inside the droplet}
\label{sec:droplet}

In this section, we analyse the flow inside the droplet, induced by the evaporative flux~\eqref{eq:Jfinal}.  The approach involves a local analysis, in which we look at the flow in the vicinity of the triple point $z_+$ (by symmetry, an analysis of the flow in the vicinity of the triple point $z_-$ will give identical results).  The starting-point is the mathematical model for the flow inside the droplet derived in Section~\ref{sec:theory}, which we now recast in terms of the mapped coordinates $(\xi,\eta)$.

We view the mapped coordinates $(\xi,\eta)$ produced by the M\"obius map~\eqref{eq:Mobius} as a set of orthogonal curvilinear coordinates.  The Euclidean metric is:
\begin{equation}
\mathd s^2=\mathd x^2+\mathd y^2.
\end{equation}
Under the conformal map:
\begin{equation}
\mathd s^2=\frac{1}{|\mathd f/\mathd z|^2}\left(\mathd\xi^2+\mathd \eta^2\right).
\end{equation}
Using the $(\rho,\phi)$ polar coordinates in the mapped space, we have:
\begin{equation}
\mathd s^2=\frac{1}{|\mathd f/\mathd z|^2}\left(\mathd\rho^2+\rho^2\mathd\phi^2\right).
\end{equation}
Therefore, we may view the $(\rho,\phi)$ polar coordinates another set of orthogonal curvilinear coordinates with scale factors:
\begin{equation}
h_\rho=\frac{1}{|\mathd f/\mathd z|},\qquad h_\phi=\frac{\rho}{|\mathd f/\mathd z|}.
\label{eq:scaleFs}
\end{equation}
In terms of the $(\rho,\phi)$ polar coordinates, the fluid velocity vector is:
\begin{equation}
\vecu=u_\rho \hatrho+u_\phi\hatphi,
\end{equation}
where
\begin{equation}
u_\rho=\hatrho\cdot\vecu,\qquad u_\phi=\hatphi\cdot\vecu.
\end{equation}
Here, the unit vectors are given by:
\begin{subequations}
\begin{eqnarray}
\hatrho&=&\cos\phi\,\hati_{\xi\eta}+\sin\phi\,\hatj_{\xi\eta},\\
\hatphi&=&-\sin\phi\,\hati_{\xi\eta}+\cos\phi\,\hatj_{\xi\eta}.
\end{eqnarray}%
\label{eq:unitRhoPhi}%
\end{subequations}%
Here, the subscript $\xi\eta$ is used to emphasize that these are Cartesian coordinates in the mapped space.

We may further write the velocity components in terms of the streamfunction $\psi$, such that:
\begin{equation}
u_\rho=\frac{1}{h_\phi}\frac{\partial\psi}{\partial\phi},\qquad u_\phi=-\frac{1}{h_\rho}\frac{\partial\psi}{\partial \rho}.
\label{eq:psiRhoPhi}
\end{equation}
In turn, the streamfunction is required to satisfy the biharmonic equation~\eqref{eq:biharmonic}, where the Laplacian is understood to mean $\nabla^2=\partial_x^2+\partial_y^2$.  In terms of the mapped-space Laplacian $\nabla_{\xi\eta}^2=\partial_\xi^2+\partial_\eta^2$, Equation~\eqref{eq:biharmonic} becomes:
\begin{equation}
\nabla_{\xi\eta}^2 \left(|\mathd f/\mathd z|^2\nabla_{\xi\eta}^2\psi\right)=0.
\end{equation}
Using the scale factors in Equation~\eqref{eq:scaleFs}, we may write this more compactly as:
\begin{equation}
\nabla_{\xi\eta}^2 \left(h_\rho^{-2}\nabla_{\xi\eta}^2\psi\right)=0.
\label{eq:biharmonicMapped}
\end{equation}
The boundary conditions for $\psi$ described in Section~\ref{sec:theory} are now represented in the mapped space as follows:
\begin{enumerate}[noitemsep]
\item No-slip on $\widetilde{F}$:  The no-slip condition $\psi_x=0$, $\psi_y=0$ no $F$ is mapped to:
\begin{equation}
\frac{\partial\psi}{\partial \rho}=0,\qquad \frac{\partial\psi}{\partial\phi}=0,\qquad \text{on }\arg(\zeta)=\alpha.
\label{eq:BCbiharmonicMapped1}
\end{equation}
\item Stress-free condition on $\widetilde{S}$: The stress-free boundary condition~\eqref{eq:noStress} is mapped to:
\begin{equation}
-\hatphi\cdot\frac{\partial\vecu}{\partial\rho}- \frac{1}{\rho}\hatrho\cdot\frac{\partial\vecu}{\partial\phi}=0,\qquad \vecu=u_\rho\hatrho+u_\phi\hatphi,\qquad \text{on }\arg(\zeta)=0.
\end{equation}
We use the properties of $\hatrho$ and $\hatphi$ in Equation~\eqref{eq:unitRhoPhi} to simplify this expression as follows:
\begin{equation}
\frac{\partial u_\phi}{\partial \rho}+\frac{1}{\rho}\frac{\partial u_\rho}{\partial \phi}-\frac{u_\phi}{\rho}=0.
\end{equation}
Using the streamfunction representation~\eqref{eq:psiRhoPhi}, this simplifies to:
\begin{equation}
\frac{\partial }{\partial \rho}\left(-\frac{1}{h_\rho}\frac{\partial\psi}{\partial\rho}\right)+\frac{1}{\rho}\frac{\partial }{\partial \phi}\left(\frac{1}{h_\phi}\frac{\partial\psi}{\partial\phi}\right)
+\frac{1}{h_\rho}\frac{1}{\rho}\frac{\partial\psi}{\partial\rho}=0.
\label{eq:BCbiharmonicMapped2}
\end{equation}
\item Mass-conservation condition:  The mass-conservation condition~\eqref{eq:massBC} is mapped to:
\begin{equation}
\frac{\partial\psi}{\partial \rho} =  \frac{Q(t)}{\beta\rho_\ell}\frac{\rho^{\frac{\pi}{2\beta}-1}  }{1+\rho^{\pi/\beta}}+\frac{2a\dot d\left(\cos\alpha-\sin\theta\right)}{\rho^2+2\rho\cos\alpha+1}.
\label{eq:BCbiharmonicMapped3}
\end{equation}
\end{enumerate}
The full calculations supporting these mappings are provided in the Supplementary Material.

\subsection{Small-$\rho$ asymptotics}

The biharmonic equation is not conformally invariant, as evidenced by mapped biharmonic equation~\eqref{eq:biharmonicMapped}, which contains the metric factor $h_\rho^{-2}$ inside the $\nabla_{\xi\eta}^2(h_\rho^{-2}\cdots)$ expression.  Consequently, the standard solution techniques available for harmonic functions are not available here.  One option is to look at solutions in terms of Goursat functions~\cite{neville2025stokes}.  In view of the complicated boundary conditions in the present application, this analysis becomes involved.  Instead, we look for analytical results in the vicinity of the triple point $z_+$, which corresponds to the limit $\rho\rightarrow 0$.  In view of the scale-factor relations~\eqref{eq:scaleFs} we may expand $h_\rho$ in the vicinity of the triple point:
hence
\begin{equation}
h_\rho\sim 2a\qquad \text{as }\rho\rightarrow 0.
\end{equation}
Therefore, in the limit of small $\rho$, the metric reduces to a flat one, and the biharmonic equation~\eqref{eq:biharmonicMapped} simplifies:
\begin{equation}
\nabla_{\xi\eta}^4\psi=0,\\
\label{eq:biharmonicMappedFlat}
\end{equation}
The boundary conditions~\eqref{eq:BCbiharmonicMapped1},~\eqref{eq:BCbiharmonicMapped2}, and~\eqref{eq:BCbiharmonicMapped3} also simplify:
\begin{enumerate}[noitemsep]
\item No-slip on $\widetilde{F}$:
\begin{equation}
\frac{\partial\psi}{\partial \rho}=0,\qquad \frac{\partial\psi}{\partial\phi}=0,\qquad \text{on }\arg(\zeta)=\alpha.
\label{eq:BCbiharmonicMappedSimple1}
\end{equation}
\item Stress-free condition on $\widetilde{S}$:
\begin{equation}
-\frac{\partial^2\psi}{\partial\rho^2}+\frac{1}{\rho^2}\frac{\partial^2\psi}{\partial\phi^2}+\frac{1}{\rho}\frac{\partial\psi}{\partial \rho}=0,\qquad \text{on }\phi=0.
\label{eq:BCbiharmonicMappedSimple2}
\end{equation}
\item Mass-conservation condition:  
\begin{equation}
\frac{\partial\psi}{\partial \rho} =  \underbrace{\frac{Q(t)}{\beta\rho_\ell} \rho^{\frac{\pi}{2\beta}-1} } +2a\dot d\left(\cos\alpha-\sin\theta\right).
\label{eq:BCbiharmonicMappedSimple3}
\end{equation}
\end{enumerate}
At this point, we recognize that the conformally-mapped bilaplacian problem~\eqref{eq:biharmonicMappedFlat}, in the vicinity of the triple point, is identical to that already worked out by Gelderblom \textit{et al.}\cite{gelderblom2012stokes}, in the context of a local wedge approximation to the droplet geometry, in the physical space.   Therefore, in the remainder of this section we only summarize the solution of~\eqref{eq:biharmonicMappedFlat}, and focus instead more on the coupling of the biharmonic equation to our newly-derived expression for the 2D evaporative flux.  
As most experiments on evaporating drops with pinned contact lines are performed at contact angles $\alpha<\pi/2$, we focus on this case here.  In this way, the dominant contribution to the boundary condition~\eqref{eq:BCbiharmonicMappedSimple3} is the underbraced one. 

Based on the prior work by Gelderblom \textit{et al.}~\cite{gelderblom2012stokes},  we propose the following solution to Equation~\eqref{eq:biharmonicMappedFlat}:
\begin{equation}
\psi=\rho^{p+1}F(\phi).
\label{eq:psiSOV}
\end{equation}
Matching powers of $p$ in Equations~\eqref{eq:BCbiharmonicMappedSimple3} and~\eqref{eq:psiSOV}, we have:
\begin{equation}
p=\frac{\pi}{2\beta}-1,
\end{equation}
where $p<0$ for $\alpha<\pi/2$.
Furthermore, we may write:
\begin{equation}
F(\phi)=c_1\cos[(p+1)\phi]+c_2\sin[(p+1)\phi]+c_3\cos[(p-1)\phi]+c_4\sin[(p-1)\phi],
\end{equation}
and where the coefficients $c_1,\cdots,c_4$ are to be determined.  The no-slip boundary conditions~\eqref{eq:BCbiharmonicMappedSimple1} give:
\begin{equation}
F(\alpha)=0,\qquad F'(\alpha)=0.
\end{equation}
The stress-free boundary condition~\eqref{eq:BCbiharmonicMappedSimple2} gives:
\begin{equation}
F''(0)+(1-p^2)F(0)=0.
\end{equation}
The mass-conservation boundary condition~\eqref{eq:BCbiharmonicMappedSimple3} gives:
\begin{equation}
(p+1)F(0)=\frac{Q(t)}{\beta\rho_\ell}.
\label{eq:FzeroDef}
\end{equation}
Combining these conditions gives a $4\times 4$ linear system for the unknown coefficients $c_1,\cdots,c_4$ (full details in the Supplementary Material).
%
The structure of the linear problem is simple, and we obtain, in the non-degenerate cases:
\begin{equation}
c_3=\tfrac{1}{2}\frac{Q(t)}{\beta\rho_\ell},\qquad c_1=-\frac{p-1}{p+1}c_3,\qquad p\neq -1,
\end{equation}
\begin{equation}
c_2=\underbrace{\left(\frac{p-1}{p+1}\right)\left[\frac{p+1-p\cos (2\alpha)+\cos (2p\alpha)}{\sin (2p\alpha)-p\sin (2\alpha)}\right]}_{=g_1(\alpha,p)}c_3,
\end{equation}
and
\begin{equation}
c_4=\underbrace{\left[\frac{p-1-p\cos(2\alpha)-\cos(2p\alpha)}{\sin (2p\alpha)-p\sin (2\alpha)}\right]}_{=g_2(\alpha,p)}c_3.
\end{equation}
This leaves the following degenerate cases:
\begin{equation}
p=-1,\qquad \sin (2p\alpha)-p\sin (2\alpha)=0,
\label{eq:degenerate}
\end{equation}
which are addressed separately below.
In the non-degenerate cases, we therefore have:
\begin{multline}
F(\phi)\\=\frac{Q(t)}{2p\beta\rho_\ell}\bigg\{-\frac{p-1}{p+1}\cos[(p+1)\phi]+g_1(\alpha,p)\sin[(p+1)\phi]+\cos[(p-1)\phi]+g_2(\alpha,p)\sin[(p-1)\phi]\bigg\}.
\end{multline}
Finally, we compute the vorticity $\omega$, in the small-$\rho$ limit.  We have:
\begin{equation}
\omega=-\frac{1}{4a^2}\nabla_{\xi\eta}^2\psi.
\end{equation}
Since terms like $\rho^{p+1}\sin[(p+1)\phi]$ and $\rho^{p+1}\cos[(p+1)\phi]$ are harmonic, we immediately get:
\begin{equation}
\omega=-\frac{1}{2a^2}\frac{Q(t)}{\beta\rho_\ell} \rho^{p-1}\bigg\{\cos[(p-1)\phi]+g_2(\alpha,p)\sin[(p-1)\phi]\bigg\}.
\end{equation}

\subsection{Degenerate cases}

We immediately rule out $p=-1$ as a degeneracy, since this corresponds to $\beta\rightarrow \infty$.  Therefore, we focus on the  zeros of $\sin (2p\alpha)-p\sin (2\alpha)$ in
Equation~\eqref{eq:degenerate}.   There are three zeros, corresponding to three degenerate cases: $\alpha=90^\circ$ ($p=0$), $\alpha\approx 133.4^\circ$, and $\alpha= 135^\circ$ ($p=1$). The  cases $\alpha=90^\circ$ and $\alpha=135^\circ$ are superficial degeneracies: by appropriate manipulation of the $4\times 4$ linear problem, a unique solution for the coefficients $c_1,\cdots,c_4$ may be found in these cases.   The degeneracy at $\alpha\approx 133.4^\circ$ is not superficial and is dealt with in detail in  Reference~\cite{gelderblom2012stokes}. All three  degenerate cases occur outside the parameter regime  of interest (specifically, $\alpha<\pi/2$), although they may be dealt with by a local (small-$\rho$) analysis similar to the one presented herein.  The degeneracy at $\alpha\approx 133.4^\circ$ may be of interest in the context of super-hydrophobic (textured) surfaces where a contact angle above $120^\circ$ is achievable~\cite{shirtcliffe2011superhydrophobicity}.

\subsection{Connection to the coffee-ring effect}

We  connect these results back to the flow observed in evaporating droplets, particularly the `coffee-ring effect' documented by Deegan~\cite{deegan1997capillary}.
Certainly, these phenomena are three-dimensional in nature.    However, we demonstrate that the same physical mechanism is at work in cylindrical droplets.  Moreover, the expressions for the evaporative-flux-driven Deegan flow in cylindrical droplets is remarkably simple, and yet has the same structure as the corresponding three-dimensional system.

We introduce the depth-averaged flow at a distance $s$ from the triple point $z_+$:
\begin{equation}
\overline{u}_{2D}(s)=\frac{1}{h}\int_0^h u(x,y')\mathd y',\qquad x=a-s.
\label{eq:us}
\end{equation}
Here, $h$ is the height of the droplet at the distance $s$ from the triple point.  The distances $s$ and $h$ and the arc-length $\ell$ are connected to the contact angle via trigonometry:
\begin{equation}
h^2+s^2=\ell^2,\qquad \tan\alpha=\frac{h}{s}.
\label{eq:trig}
\end{equation}
Equation~\eqref{eq:us} may be connected to the streamfunction via:
\begin{equation}
\overline{u}_{2D}(s)=\frac{1}{h}\left[\psi(x=a-s,h)-\psi(x=a-s,0)\right].
\end{equation}
Since $y=h$ corresponds to the free surface and $y=0$ corresponds to the flat interface, this simplifies to:
\begin{equation}
\overline{u}(s)=\frac{1}{\ell \sin\alpha } \left(\frac{\ell}{2a}\right)^{p+1}\left[F(0)-\cancel{F(\alpha)}\right],\qquad \text{as }\ell\rightarrow 0.
\end{equation}
Here, we have used $\rho\sim \ell/(2a)$, valid in the limit as $s\rightarrow 0$ (\textit{cf}. Equation~\eqref{eq:rhoLim}).
Referring back to the expression for the expression~\eqref{eq:FzeroDef} for $F(0)$ valid for $p<1$ ($\alpha<\pi/2$), this gives:
\begin{equation}
\overline{u}_{2D}(s)=\frac{Q(t)}{2a(p+1)\beta\rho_\ell\sin\alpha }\left(\frac{\ell }{2a}\right)^{p},
\label{eq:uDeegan2}
\end{equation}
where $\ell$ and $s$ are connected via Equation~\eqref{eq:trig}.

Equation~\eqref{eq:uDeegan2} may also be derived using a simple mass-conservation argument.  We use the right-angled triangle defined by Equation~\eqref{eq:trig} as a control area.  Let the mass of the fluid inside this control area by $M_s$.  We have:
\begin{equation}
\frac{\mathd M_s}{\mathd t}=\int_0^\ell \myJ\mathd \ell.
\end{equation}
Using Equation~\eqref{eq:Jlim2D}, this becomes:
\begin{equation}
\frac{\mathd M_s}{\mathd t}=\frac{Q(t)}{2a\beta}\frac{1}{(2a)^p}\frac{\ell^{p+1}}{p+1}.
\end{equation}
This mass loss is balanced by fluid entering the control volume through the side of the triangle of length $h$.  Therefore:
\begin{equation}
\frac{\mathd M_s}{\mathd t}=\rho_\ell \overline{u}_{2D}(s) h.
\end{equation}
We equate these equivalent expressions  to get an expression for $\overline{u}_{2D}(s)$ which agrees exactly with the previously-derived Equation~\eqref{eq:uDeegan2}.

A similar mass-balance approach may be pursued in three dimensions.  We use a control volume as opposed to a control area. Looking down on the 3D spherical-cap droplet from above, the projection of the control volume on to the plane of the substrate is an annulus with inner radius $a-s$ and outer radius $a$.   The area of the annulus is:
\begin{equation}
A_{s,p}=\pi a^2-\pi (a-s)^2=2\pi a s + O(s^2).
\end{equation}
Correcting for the projection gives the area of the control volume in contact with the atmosphere
\begin{equation}
A_{s}=2\pi a\frac{s}{\cos\alpha}.
\end{equation}
Hence:
\begin{equation}
\mathd A_s=2\pi a\frac{\mathd s}{\cos\alpha}.
\end{equation}
The mass loss from the control volume is therefore:
\begin{equation}
\frac{\mathd M_s}{\mathd t}=\int_0^s J_{3D}\mathd A_s
\end{equation}
Using Equation~\eqref{eq:J3D}, this may be evaluated to give:
\begin{equation}
\frac{\mathd M_s}{\mathd t}=\frac{2\pi a J_0}{a^p(p+1)\cos\alpha}s^{p+1},
\end{equation}
where $J_0$ is a prefactor.
This is balanced by a mass flux across an area $2\pi a s\tan\alpha$:
\begin{equation}
\frac{\mathd M_s}{\mathd t}=\rho_\ell \overline{u}_{3D}(s)2\pi a s\tan\alpha
\end{equation}
(here, $\rho_\ell$ is a mass per unit volume).  Matching the two expressions for $\mathd M_s/\mathd t$ gives:
\begin{equation}
\overline{u}_{3D}(s)=\frac{J_0}{(p+1)}\frac{1}{\rho_\ell\sin\alpha}(s/a)^p.
\label{eq:uDeegan3}
\end{equation}
Comparing Equations~\eqref{eq:uDeegan2} and~\eqref{eq:uDeegan3}, it can be seen that the scaling exponent is the same for $\overline{u}_{2D}$ and $\overline{u}_{3D}$, namely $\overline{u}\sim s^p$ as $s\rightarrow 0$, giving the radially-outward Deegan flow.

\section{Discussion and Conclusions}
\label{sec:conc}

We have formulated a mathematical description of droplet evaporation in two dimensions (the `cylindrical cap geometry').  The description fully accounts for the vapour diffusion in the atmosphere and the flow inside the droplet, and applies in the diffusion-limited regime in the atmosphere and the Stokes-flow regime inside the droplet.
We have solved for the vapour concentration in the atmosphere by extending Yariv's theory~\cite{yariv2023lifetime} to finite contact angles.  In doing so, we have recovered the behaviour $\myJ\sim (\ell/a)^{\lambda-1}$ in the vicinity of the triple point, where $\ell$ is the distance in arc-length from the triple point, and $\lambda=  \pi/[2(\pi-\alpha)]$.  Thus, the local behaviour of the flux in the vicinity of the triple point depends  only on the wedge-shape geometry, which is the same in 2D and 3D.  Our results on the vapour-concentration modelling also include a closed-form expression for the integral flux, $F_{2D}=\int \myJ\mathd \ell$.  Our results are independent of an (arbitrary) far-field boundary location and therefore build upon previous work in the field.

We have  formulated the problem of the Stokes flow inside the droplet in terms of the biharmonic equation, with a mass-conservation boundary condition coupled to $\myJ$.  To simplify the mathematics, the atmosphere and droplet problems are formulated in  conformally-mapped domains, which take the form of contiguous wedges.  
The conformally-mapped biharmonic equation contains metric terms, which make exact analytical solutions difficult.  While we do not rule out developing such solutions in future work, for the present purposes, we have computed explicit solutions valid locally in the vicinity of the triple point.  The form of these solutions mirrors the local analysis previously carried out by Gelderblom \textit{et al.}~\cite{gelderblom2012stokes}; although here we emphasize the unity of the present approach, in terms of the coupling of the two problems (atmosphere and droplet), and the conformal mapping of these problems into two contiguous wedge geometries. 


Although 2D cylindrical-cap droplets are not physical entities, they are extensively studied in the literature for the theoretical understanding they give to the corresponding 3D spherical-cap analogue.  The 2D formulation provides a useful reduced model of evaporating droplets, provided that the distinction between local and global behaviour is kept in mind. 
Globally, the 2D and 3D problems differ, as the corresponding evaporation rates have different functional forms. Locally, however, the 2D and 3D problems share the same wedge structure at the triple point, and the present analysis shows that this correspondence extends from the evaporative-flux singularity to the scaling of the induced internal flow. The usefulness of the of the 2D formulation is therefore not as a quantitative representation of a 3D droplet, but as a tractable model that illustrates the local coupling between evaporation and flow at the triple point.

\subsection*{Declaration of AI use}

OpenAI, GPT-5.6 was used in the development of the conformal map and the matched asymptotic expansions in this work.  All results were verified by hand, step-by-step, and the resulting calculations are included in the Supplementary Material.  
No artificial intelligence was used in the writing of the manuscript.
     
\subsection*{Acknowledgments}

This publication has emanated from research supported in part by a grant from Research Ireland under Grant Number 18/CRT/6049. 
Helpful discussions with M. D. Bustamante and A. Amirfazli are acknowledged.  The collaboration between the authors and A. Amirfazli was supported by the UCD School of Mathematics and Statistics Distinguished Visitor Programme (2026).

\newpage

\section*{Supplementary Material -- Overview}

\setcounter{section}{0}
\setcounter{subsection}{0}
\setcounter{equation}{0}
\setcounter{figure}{0}
\setcounter{table}{0}
\renewcommand{\theequation}{S\arabic{equation}}

This document contains additional detailed calculations in support of the results stated in the main manuscript.  The following detailed calculations are presented.  Note: \textit{Enumerated items match directly with section headings.  This introductory section is un-numbered.}
\begin{enumerate}[noitemsep]
\item Detailed description of the conformal map, including the mapping of the boundary faces $F_\pm$ and $S$;
\item Matched asymptotic expansions: Solution of Laplace's equation in the inner region;
\item Matched asymptotic expansions: Matching inner and outer solutions;
\item Evaluation of integral flux in the large-$t$ limit;
\item The Stokes-flow problem in the mapped domain, including:
    \begin{itemize}[noitemsep]
		\item  Incompressibility of flow inside droplet;
		\item Rigorous derivation of mass-balance equations across the free surface $S$;
		\item Mathematical modelling of CR and CCA modes of evaporation.
		\item Geometry of the conformal map, streamfunction and boundary-condition mapping;
		\item Stokes Flow in the mapped domain;
		\end{itemize}
\item Reduction of Stokes flow to $4\times4$ linear problem.
\end{enumerate}
We refer to the figures in the main manuscript instead of reproducing them here.  However, in order to make the supplementary material as self-contained as possible, we repeat the basic mathematical model here below.
\newpage

\section*{Basic Mathematical Model}

\subsection{The model inside the droplet}
The vapour concentration is  described by the diffusion equation:
\begin{subequations}
\begin{equation}
\frac{\partial c}{\partial t}=D_v\nabla^2 c,\qquad \text{in }\Omega_a,\qquad t>0,
\end{equation}
where $D_v$ is the vapour diffusivity.  At the interface between the substrate and the atmosphere (faces $F_-$ and $F_+$), Neumann boundary conditions apply:
\begin{equation}
\frac{\partial c}{\partial y}=0,\qquad \text{on }F_+\text{ and }F_-.
\end{equation}
On the free surface $S$ separating the droplet and the atmosphere, the vapour concentration is taken as saturated:
\begin{equation}
c=c_s\qquad \text{on }S.
\end{equation}
Finally, the far-field boundary condition is:
\begin{equation}
\lim_{r\rightarrow \infty}c=c_\infty.
\end{equation}
Here, we have introduced the radial coordinate in the $xy$ plane, %
\begin{equation}
r=\sqrt{x^2+y^2}.
\end{equation}
The initial condition is given by:
\begin{equation}
c=c_\infty,\qquad t=0.
\end{equation}%
The evaporative flux from the droplet into the atmosphere is given by:
\begin{equation}
J=-D_v\hatn\cdot\nabla c,\qquad \text{on }S,
\label{eq:Jdef}
\end{equation}
where $\hatn$ is a normal vector to the surface $S$, pointing from the liquid into the atmosphere.
\label{eq:model_all}%
\end{subequations}%

\subsection{The model inside the droplet}

The flow inside the droplet is characterized by its low Reynolds number and consequently, may be described using the Stokes equation:
\begin{equation}
-\nabla p+\mu \nabla^2\vecu=0,\qquad \nabla\cdot\vecu=0,
\label{eq:stokes}
\end{equation}
where $p$ is pressure and $\mu$ is the dynamic viscosity.
Boundary conditions are: no-slip at the liquid-solid interface, and a stress-free condition on the surface $S$:
\begin{subequations}
\begin{equation}
\widehat{\bm{t}}\cdot\left(\nabla \vecu+\nabla\vecu^T\right)\cdot\hatn=0,\qquad\text{on }S.
\label{eq:noStress}
\end{equation}
Here, $\hatt$ is any unit tangent vector to the free surface $S$.
The velocity field $\vecu$ is coupled to the evaporation flux via the relation:
\begin{equation}
V_n+\frac{J}{\rho_\ell}=\hatn\cdot\vecu,\qquad\text{on }S,
\label{eq:massBC}
\end{equation}%
\label{eq:StokesBCs}%
\end{subequations}%
where $\rho_{\ell}$ is the density of the liquid inside the droplet.  Here, $V_n$ is the normal velocity of the surface $S$, which is non-zero for an evaporating droplet.

The flow field $\vecu$ is assumed to be incompressible, $\nabla\cdot\vecu=0$.  Consequently, $\vecu$ may be described in terms of a streamfunction:
\begin{equation}
u=\frac{\partial \psi}{\partial y},\qquad v=-\frac{\partial\psi}{\partial x}.
\end{equation}
Under these conditions,  Equation~\eqref{eq:stokes} reduces to the biharmonic equation:
\begin{equation}
\nabla^4\psi=0,\qquad \text{in }\Omega_d.
\label{eq:biharmonic}
\end{equation}
The boundary conditions on $\vecu$ are no-slip on the substrate face $F$, which equates to:
\begin{equation}
\psi_x=0,\qquad \psi_y=0,\qquad \text{on }F,
\end{equation}
as well as the previously-introduced stress-free boundary condition~\eqref{eq:noStress}.
 To close the system of equations, a final boundary condition is required.  This is obtained by imposing the conservation of mass across the free surface $S$, the precise statement of which is given in Equation~\eqref{eq:massBC}.

\section{Conformal map}

We refer to the geometry outlined in the main manuscript in Figures 1--2.  We introduce complex coordinates for the points where the chord $y=d$ intersects the circle.  These are:
\begin{equation}
z_-=-a+\imag d,\qquad z_+=a+\imag d,
\end{equation}
or, in terms of polar coordinates:
\begin{equation}
z_-=R\mathe^{\imag \theta_-},\qquad z_+=R\mathe^{\imag\theta_+},
\end{equation}
where:
\begin{equation}
\theta_-=\tfrac{1}{2}\pi+\alpha,\qquad \theta_+=\tfrac{1}{2}\pi-\alpha.
\end{equation}
Notice:
\begin{equation}
z_+-z_-=2R\sin\alpha.
\label{eq:zpmDiff}
\end{equation}
We define the M\"obius transformation:
\begin{equation}
\zeta=f(z),\qquad f(z)=-\mathe^{\imag\alpha}\left(\frac{z-z_+}{z-z_-}\right).
\label{eq:mobius}
\end{equation}
\begin{equation}
z=f^{-1}(\zeta),\qquad f^{-1}(\zeta)=\frac{\zeta z_-+\mathe^{\imag\alpha}z_+}{\zeta+\mathe^{\imag\alpha}}.
\label{eq:inverseT}
\end{equation}
We write:
\begin{equation}
\zeta=\xi+\imag\eta.
\end{equation}
By direct calculation we obtain:
\begin{subequations}
\begin{equation}
\xi= \frac{2a(y-d)\sin\alpha            -\left[ x^2-a^2+(y-d)^2\right]\cos\alpha }{  (x+a)^2+(y-d)^2}.
\end{equation}
and
\begin{equation}
\eta=-\frac{\left[ x^2-a^2+(y-d)^2\right]\sin \alpha +2a(y-d)\cos\alpha}{(x+a)^2+(y-d)^2}.
\end{equation}
\label{eq:xieta}
\end{subequations}
Other useful expressions include the following.  The derivative of the map is given by:
\begin{equation}
\frac{\mathd f}{\mathd z}=-\mathe^{\imag\alpha}\frac{z_+-z_-}{(z-z_-)^2}.
\label{eq:dfdz}%
\end{equation}
The derivative of the inverse map is given by:
\begin{equation}
\frac{\mathd f^{-1}}{\mathd \zeta}=-\mathe^{\imag\alpha}\frac{z_+-z_-}{(\zeta-\mathe^{\imag\alpha})^2}.
\label{eq:dfInvdzeta}%
\end{equation}
We also use polar coordinates in the mapped space, with $\rho=\sqrt{\xi^2+\eta^2}$ and $\phi$ the polar angle starting from the $\xi$-axis and in tending in the direction of  increasing $\eta$.  In this context, we use the identity:
\begin{equation}
\frac{\mathd f^{-1}}{\mathd \zeta}\frac{\mathd f}{\mathd z}=1
\end{equation}
together with Equation~\eqref{eq:dfInvdzeta} to write:
\begin{equation}
\left|\frac{\mathd f}{\mathd z}\right|=\frac{\rho^2+2\rho\cos(\phi-\alpha)+1}{2a}.
\label{eq:dfdzMetric}
\end{equation}
Equation~\eqref{eq:dfdzMetric} will be particularly useful as it represents a metric factor used to transform between different orthogonal coordinate systems.

Under the conformal  map~\eqref{eq:mobius}, the domain $\Omega$ gets mapped to $\widetilde{\Omega}$:
\begin{equation}
\widetilde{\Omega}=f(\Omega).
\end{equation}
Correspondingly, the boundaries $F\pm$ and $S$ get mapped to $\widetilde{F_{\pm}}$ and $\widetilde{S}$.  We are interested in the shape of the mapped boundaries:
\begin{enumerate}
\item $F_-$:  We have $z=x+\imag d$, hence
\begin{equation}
\zeta=-\mathe^{\imag\alpha}\left[ \frac{x+\imag d-(a+\imag d)}{x+\imag d-(-a+\imag d)}\right]=\mathe^{\imag \alpha}\left(\frac{a-x}{a+x}\right).
\label{eq:zetaF}
\end{equation}
For $F_-$, we have $x<-a$, so $F_-$ gets mapped to $\zeta=(-\infty,-1)\mathe^{\imag\alpha}$.  The droplet extremity $z_-=(-a,d)$ gets mapped to $-\infty\mathe^{\imag\alpha}$, and the far-field point $x\rightarrow -\infty$ gets mapped to $-\mathe^{\imag\alpha}$.

\item $F_+$: Equation~\eqref{eq:zetaF} still holds, with $x>a$.  Therefore, $F_+$ gets mapped to $\zeta=(-1,0)\mathe^{\imag\alpha}$.  The droplet extremity $(a,d)$ gets mapped to the origin, and the far-field point $x\rightarrow \infty$ gets mapped to $-\mathe^{\imag\alpha}$.

\end{enumerate}

\noindent Overall, the faces $F_{\pm}$ get mapped to the ray:
\[
\arg(\zeta)=\pi+\alpha.
\]
The far-field points $|x|\rightarrow\infty$  both get mapped to the point $\mathe^{\imag\alpha}$ on this ray.

\begin{enumerate}[resume]

\item Boundary face $S$: On the curved face $S$, we have $z=R\mathe^{\imag\theta}$, where $\theta_+<\theta<\theta_-$ (we use $\theta$ as the polar coordinate).  Hence:
\begin{eqnarray*}
\zeta&=&-\mathe^{\imag\alpha} \left(\frac{R\mathe^{\imag\theta}-R\mathe^{\imag\theta_+}}{R\mathe^{\imag\theta}-R\mathe^{\imag \theta_-}}\right),\\
     &=&-\mathe^{\imag\alpha} \left(\frac{\mathe^{\imag\theta}-\mathe^{\imag\theta_+}}{\mathe^{\imag\theta}-\mathe^{\imag \theta_-}}\right).
\end{eqnarray*}
We use a trigonometric identity:
\begin{equation}
\mathe^{\imag A}-\mathe^{\imag B}=2\imag \mathe^{\imag (A+B)/2}\sin\left(\frac{A-B}{2}\right).
\end{equation}
Hence:
\begin{equation}
\zeta=\xi+\imag\eta=-\frac{\sin\left(\frac{\theta-\theta_+}{2}\right)}{\sin\left(\frac{\theta-\theta_-}{2}\right)}+0\imag.
\label{eq:rhoTheta}
\end{equation}
Hence, $\eta=0$ on $S$.  We identify:
\begin{equation}
\mathcal{R}(\theta)=-\frac{\sin\left(\frac{\theta-\theta_+}{2}\right)}{\sin\left(\frac{\theta-\theta_-}{2}\right)},\qquad \theta_+<\theta<\theta_-.
\end{equation}
We may check some distinguished points:
\begin{itemize}[noitemsep]
\item $\theta=\theta_+$, hence $\xi=0$.  So $z_+$ gets mapped to $\xi=0$.
\item $\theta=\theta_-$, hence $\xi=\infty$.  So $z_-$ gets mapped to $\xi=\infty$.
\end{itemize}
Therefore, it is important to note that $z_-$ gets mapped both to $-\infty\mathe^{\imag\alpha}$ and $\infty$ -- it depends on the direction of approach to $z_-$:
\[
\lim_{\stackrel{z\rightarrow z_-}{z\in F_-}}f(z)=-\infty \mathe^{\imag\alpha},\qquad 
\lim_{\stackrel{z\rightarrow z_-}{z\in S}}f(z)=-\infty .
\]
In any case, the curved face $S$ gets mapped to:
\[
\arg(\zeta)=0,
\]
or, equivalently, 
\[
\arg(\zeta)=2\pi.
\]
\end{enumerate}
So overall, the M\"obius map takes the atmospheric domain $\Omega$ to the wedge:
\begin{equation}
\widetilde{\Omega}=\{\zeta\in \mathbb{C}|\alpha+\pi<\arg(\zeta)<2\pi \}.
\end{equation}
The mapped domain is shown schematically in Figure~3 of the main paper.

\subsection{Mapping of the Boundary Condition -- Flat Face}

We are interested in the behavior of the boundary condition $(\partial c/\partial y)_{F_\pm}=0$  under the conformal map.  We have:
\begin{eqnarray*}
\left(\begin{array}{c}c_x\\ c_y\end{array}\right)&=&\left(\begin{array}{cc} \xi_x & \eta_x \\ \xi_y & \eta_y\end{array}\right)\left(\begin{array}{c}c_\xi\\c_\eta\end{array}\right),\\ 
                                                    &=& \mathbb{J}\left(\begin{array}{c}c_\xi\\c_\eta\end{array}\right),
\end{eqnarray*} 
where we introduce the notation $\mathbb{J}$ for the Jacobian matrix.  We have:
\begin{eqnarray*}
\mathbb{J}\mathbb{J}^T&=& \left(\begin{array}{cc} \xi_x & \eta_x \\ \xi_y & \eta_y\end{array}\right)\left(\begin{array}{cc} \xi_x & \xi_y \\ \eta_x & \eta_y\end{array}\right),\\
    &=&\left(\begin{array}{cc} \xi_x+\eta_x^2&\xi_x\xi_y+\eta_x\eta_y\\ \xi_y\xi_x+\eta_y\eta_x & \xi_y^2+\eta_y^2\end{array}\right),\\
		&\stackrel{\text{CR}}{=}&\left(\begin{array}{cc}\xi_x^2+\eta_x^2&0\\0&\xi_y^2+\eta_y^2\end{array}\right),\\
		&\stackrel{\text{CR}}{=}&\left|\frac{\mathd f}{\mathd z}\right|^2\left(\begin{array}{cc}1&0\\0&1\end{array}\right).
\end{eqnarray*}
Here, `CR' indicates that the Cauchy--Riemann (CR) conditions have been applied.
It follows that:
\[
\mathbb{J}=\left|\frac{\mathd f}{\mathd z}\right|\underbrace{\left(\begin{array}{cc}\cos\Phi&-\sin\Phi\\\sin\Phi&\cos\Phi\end{array}\right)}_{=\mathbb{Q}},
\]
where $\Phi$ is the (local) angle which parametrizes the orthogonal matrix $\mathbb{Q}$.  

We now revisit the boundary condition $(\partial c/\partial y)_{F_\pm}=0$.  This can be re-written as %
\[
(-\hatj_{xy})\cdot\nabla_{xy}c=0,\qquad (x,y)\in F_\pm.
\]  
Here, we are using $\hatj_{xy}$ to denote the unit vector in the $y$-direction in the physical space, and similarly, we are using $\nabla_{xy}u$ to denote the gradient in the physical space, hence:
\[
\nabla_{xy}c=\left(\begin{array}{c}c_x\\c_y\end{array}\right).
\]
Similarly, we use $\nabla_{\xi\eta}c$ to denote the gradient in the mapped space:
\[
\nabla_{\xi\eta}u=\left(\begin{array}{c}c_\xi\\c_\eta\end{array}\right).
\]
We also use angle brackets to denote the $L^2$ pairing:
\[
(-\hatj_{xy})\cdot\nabla_{xy}c=\langle (-\hatj_{xy}),\nabla_{xy}c\rangle.
\]
We have:
\begin{eqnarray*}
(-\hatj_{xy})\cdot\nabla_{xy} c&=&\langle (-\hatj_{xy}),\nabla_{xy}c\rangle,\\
                 &=&\langle (-\hatj_{xy}),\left|\frac{\mathd f}{\mathd z}\right|\mathbb{Q} \nabla_{\xi\eta}c\rangle,\\
								 &=&\left|\frac{\mathd f}{\mathd z}\right|\langle \mathbb{Q}^T(-\hatj_{xy}),\nabla_{\xi\eta}c\rangle.
\end{eqnarray*}
Here, $\mathbb{Q}^T(-\hatj_{xy})$ is the rotation of the pertinent outward-pointing unit normal under the conformal mapping.  Since the faces $F_{\pm}$ with outward pointing unit normal $-\hatj_{xy}$ gets mapped to the ray $\arg(\zeta)=\alpha$, it follows that:
\[
\mathbb{Q}^T(-\hatj_{xy})=(-\sin\alpha,\cos\alpha).
\]
Hence, the mapped boundary condition is:
\begin{equation}
-\sin\alpha\frac{\partial c}{\partial\xi}+\cos\alpha\frac{\partial c}{\partial \eta}=0,\qquad \xi=\rho(x)\cos\alpha,\qquad \eta=\rho(x)\sin\alpha.
\label{eq:mappedFlat}
\end{equation}

\subsection{Mapping of the Boundary Condition -- Curved  Face}

As the boundary condition on the curved face is a simple Dirichlet condition, we have:
\[
c(\xi,\eta)=c_s,\qquad \xi>0,\qquad \eta=0.
\]

In what follows, we will be interested in understanding the value of the evaporative flux, defined previously in Equation~\eqref{eq:Jdef}.
For this purpose, we work out the flux $J$ in the mapped space.  Using the same logic as before, we have:
\begin{eqnarray*}
\hatn\cdot\nabla_{xy} c&=&\langle \hatn,\nabla_{xy}c\rangle,\\
                 &=&\langle \hatn,\left|\frac{\mathd f}{\mathd z}\right|\mathbb{Q} \nabla_{\xi\eta}c\rangle,\\
								 &=&\left|\frac{\mathd f}{\mathd z}\right|\langle \mathbb{Q}^T\hatn,\nabla_{\xi\eta}c\rangle.
\end{eqnarray*}
Here, $\mathbb{Q}^T\hatn$ is the rotation of the unit vector $\hatn$ under the conformal mapping.  This gives:
\[
\mathbb{Q}^T\hatn=-\hatj_{\xi\eta},
\]
where $\hatj_{\xi\eta}$ is unit vector in the $\eta$-direction in the mapped space.  Hence:
\begin{equation}
J=-D_v\left(\hatn\cdot\nabla_{xy} c\right)_{r=R}=D_v\left|\frac{\mathd f}{\mathd z}\right|_{\eta=0}\left(\frac{\partial c}{\partial\eta}\right)_{\eta=0}.
\label{eq:mapped1}
\end{equation}
This is the required transformation law of the directional derivative of $c$ on the curved boundary $S$.

We have:
\begin{subequations}
\begin{eqnarray}
\frac{\mathd f}{\mathd z}&=&\frac{\mathd}{\mathd z}\left[-\mathe^{\imag\alpha}\frac{z-z_+}{z-z_-}\right],\\
                           &=&-\mathe^{\imag\alpha}\frac{z_+-z_-}{(z-z_-)^2}.
\end{eqnarray}%
\label{eq:dfdz}%
\end{subequations}%
Referring back to Equation~\eqref{eq:inverseT}, we have:
\begin{eqnarray*}
z-z_-&=&\frac{\zeta z_-+\mathe^{\imag\alpha}z_+}{\zeta+\mathe^{\imag\alpha}}-z_-,\\
     &=&\frac{\zeta z_-+\mathe^{\imag\alpha}z_+-z_-\left(\zeta+\mathe^{\imag\alpha}\right)}{\zeta+\mathe^{\imag\alpha}},\\
		 &=&\frac{\cancel{\zeta z_-}+\mathe^{\imag\alpha}z_+-\cancel{\zeta z_-}-\mathe^{\imag\alpha}z_-}{\zeta+\mathe^{\imag\alpha}},\\
\end{eqnarray*}
Hence:
\begin{equation}
z-z_-=\frac{\mathe^{\imag\alpha}(z_+-z_-)}{\zeta+\mathe^{\imag\alpha}}.
\label{eq:zzm}
\end{equation}
Putting it all together, we have:
\begin{eqnarray*}
\frac{\mathd f}{\mathd z}&\stackrel{\text{Eq.~\eqref{eq:dfdz}}}{=}&-\mathe^{\imag\alpha}\frac{z_+-z_-}{(z-z_-)^2},\\
                         &\stackrel{\text{Eq.~\eqref{eq:zzm}}}{=}& \frac{-\cancel{\mathe^{\imag\alpha}}\cancel{(z_+-z_-)}}{ \cancel{\mathe^{\imag\alpha}}\cancel{(z_+-z_-)}\mathe^{\imag\alpha}(z_+-z_-)}(\zeta+\mathe^{\imag\alpha})^2,\\
												&\stackrel{\text{Eq.~\eqref{eq:zpmDiff}}}{=}&-\mathe^{-\imag\alpha}\frac{(\zeta+\mathe^{\imag\alpha})^2}{2R\sin\alpha}.
\end{eqnarray*}
The final equation here will be of use later, so we make special note of it here:
\begin{equation}
\frac{\mathd f}{\mathd z}=-\mathe^{-\imag\alpha}\frac{(\zeta+\mathe^{\imag\alpha})^2}{2R\sin\alpha}.
\label{eq:dfdz}
\end{equation}
Furthermore, if we introduce $(\rho,\phi)$ as polar coordinates in the mapped space, we may write:
\begin{equation}
\left|\frac{\mathd f}{\mathd z}\right|= \frac{\rho^2+2\rho\cos\alpha+1}{2R\sin\alpha},
\label{eq:dfdz1}
\end{equation}
or:
\begin{equation}
\left|\frac{\mathd f}{\mathd z}\right|= \frac{\rho^2+2\rho\cos\alpha+1}{2a}.
\label{eq:dfdz2}
\end{equation}
Putting it all together, we have:
\begin{equation}
J=D_v\frac{|\zeta+\mathe^{\imag\alpha}|^2}{2R\sin\alpha}\left(\frac{\partial c}{\partial \eta}\right)_{\eta=0},
\end{equation}
valid on $\zeta=\xi+0\imag$.  Hence:
\begin{equation}
J=D_v\frac{|\zeta+\mathe^{\imag\alpha}|^2}{2R\sin\alpha}\left(\frac{\partial c}{\partial \eta}\right)_{\eta=0}.
\label{eq:j1}
\end{equation}

\subsection{Summary of problem in the mapped space}

We summarize the problem in the mapped space $\widetilde{\Omega}$.  The mapped diffusion equation reads:
\begin{subequations}
\begin{equation}
\frac{\partial c}{\partial t}=D_v|f'(z)|^2\nabla_\zeta^2c,\qquad \text{in } \widetilde{\Omega},\qquad t>0,
\end{equation}
The boundary conditions ($t>0$)  on the mapped faces are:
\begin{eqnarray}
c&=&c_\infty,\qquad t=0,\\
-\sin\alpha\frac{\partial c}{\partial\xi}+\cos\alpha\frac{\partial c}{\partial \eta}&=&0,\qquad \text{on }\widetilde{F},\\
c&=&c_s,\qquad\text{on }\widetilde{S}.
\end{eqnarray}
The far-field boundary conditions ($t>0$) are:
\begin{equation}
\lim_{\stackrel{\xi\rightarrow\infty}{\eta=0}}c=c_\infty,
\end{equation}%
and
\begin{equation}
\lim_{z\rightarrow -\mathe^{\imag\alpha} }c=c_\infty.
\end{equation}%
Finally, the initial condition reads:
\begin{equation}
c=c_\infty,\qquad t=0.
\end{equation}
\label{eq:summaryMapped}%
\end{subequations}%

\section{Laplace Equation -- Inner Solution}
\label{sec:inner}

In the inner region, close to the droplet surace, we may assume that the vapour concentration has reached equilibrium, meaning the heat equation reduces to Laplace's equation.  As Laplace's equation is invariant under conformal mapping, the mathematical model in the mapped coordinates reduces to:
\begin{equation}
\nabla_{\xi\eta}^2 c_{inner}=0.
\end{equation}
Here, we use the subscript `inner' to denote a solution valid in the inner region, near the droplet surface.

We look at the inner solution $\nabla_{\xi\eta}^2 c_{inner}$=0, on the wedge $\alpha+\pi<\arg(\zeta)<2\pi$.  If instead, we take $\varphi=\arg(\zeta)-(\alpha+\pi)$, we obtain the geometry in Figure~4 of the main paper (`modified mapped domain').   Laplace's equation in the wedge-shaped domain in  that figure can be solved using separation of variables.  We use polar coordinates, with a polar angle $\varphi$ and a radial distance $\rho$.  Thus, the wedge angle is $\varphi=\beta$, where $\beta=\pi-\alpha$.  To absorb the inhomogeneous boundary condition along the ray $\varphi=\beta$, we use the trial solution $c=c_s+v$.  The equation to be solved reads:
\begin{subequations}
\begin{eqnarray}
\nabla^2 v&=&0,\\
v&=&0,\qquad \varphi=\beta,\\
\frac{\partial v}{\partial \varphi}&=&0,\qquad \varphi=0.
\end{eqnarray}
Laplace's equation in an infinite wedge does not have a unique solution, as there are not enough boundary conditions to constrain all of the constants of integration~\cite{strauss2007partial}.  To obtain a unique solution, we apply the extra condition:
\begin{equation}
v\sim - C\ln |\rho-1|,\qquad \text{as }\rho\rightarrow 1,
\label{eq:lnsln}
\end{equation}
where $C$ is a free (positive) constant.
Condition~\eqref{eq:lnsln} ensures that the complete inner solution matches up with the outer solution in the overlap region (boundary faces).
\label{eq:laplaceWedge}
\end{subequations}
We solve Equation~\eqref{eq:laplaceWedge} using separation of variables, $v=F(\rho)\Phi(\phi)$.  This yields $\Phi''+\lambda^2\Phi=0$.  The boundary condition $\Phi'(0)=0$ implies $\Phi\propto \cos(\lambda\phi)$.  The boundary condition $\Phi(\beta)=0$ fixes $\lambda$ as:
\[
\lambda \beta=\left(n+\tfrac{1}{2}\right)\pi,\qquad n\in \{0,1,2,\cdots\}.
\]
Hence,
\[
\lambda=\lambda_n,\qquad \lambda_n:=\frac{\left(n+\tfrac{1}{2}\right)\pi}{\beta},
\]
or
\[
\lambda_n=\frac{\left(n+\tfrac{1}{2}\right)\pi}{\pi-\alpha},
\]
where $n\in \{0,1,2,\cdots\}$.  Consequently, the angular part of the solution reads:
\[
\Phi_n(\varphi)=\cos\left(\frac{\left(n+\tfrac{1}{2}\right)\pi}{\pi-\alpha}\varphi\right).
\]
We revisit the radial part of the solution, which by separation of variables looks like:
\[
\rho^2\left[\frac{\frac{1}{\rho}\frac{\mathd}{\mathd \rho}\left(\rho\frac{\mathd F_n}{\mathd \rho}\right)}{F_n}\right]=-\frac{\Phi_n''}{\Phi_n}=\lambda_n^2,
\]
Hence:
\[
\frac{\mathd^2F_n}{\mathd \rho^2}+\frac{1}{\rho}\frac{\mathd F_n}{\mathd \rho}-\frac{\lambda_n^2}{\rho^2}F_n=0,
\]
with solution:
\[
F_n=\lambda_n \rho^{\lambda_n}+B_n\rho^{-\lambda_n},
\]
where $A_n$ and $B_n$ are constants of integration.
We define:
\[
F_n(\rho)=\begin{cases}\rho^{\lambda_n},&0<\rho\leq 1,\\
                       \rho^{-\lambda_n},&\rho\geq 1.\end{cases} 
\]
A solution of Laplace's equation in the wedge is therefore:
\[
\sum_{n=0}^\infty g_n \cos(\lambda_n\varphi)F_n(\rho),
\]
where the $g_n$ are constants of integration.  We have the following important theorem:

\begin{theorem}
\label{thm:Glog}
The solution of Equation~\eqref{eq:laplaceWedge} satisfying all the boundary conditions including the behavioural boundary condition~\eqref{eq:lnsln} is:
\[
G(\rho,\varphi)=\sum_{n=0}^\infty \frac{1}{\lambda_n}\cos(\lambda_n \varphi)F_n(\rho).
\]
\end{theorem}

\begin{proof} In {\textbf{Case 1}}, we take $\varphi=0$ and $\rho<1$.  We have:
\begin{eqnarray*}
G(\rho,0)&=&\sum_{n=0}^\infty \frac{1}{ \left(n+\tfrac{1}{2}\right)\frac{\pi}{\beta}}\rho^{\left(n+\tfrac{1}{2}\right)\frac{\pi}{\beta}},\\
      &=&\frac{2\beta}{\pi}\sum_{n=0}^\infty \frac{1}{2n+1} \rho^{\pi/2\beta}\rho^{(\pi/2\beta)2n},\\
			&=&\frac{2\beta}{\pi}\sum_{n=0}^\infty \frac{1}{2n+1} {\rho^{\pi/(2\beta)}}^{2n+1},\\
			&\stackrel{\rho<1}{=}&\frac{2\beta}{\pi}\tanh^{-1}(\rho^{\pi/(2\beta)}),\\
			&=&\frac{\beta}{\pi}\ln \left(\frac{1+\rho^{\pi/(2\beta)}}{1-\rho^{\pi/(2\beta)}}\right).
\end{eqnarray*}
Since we are looking at the limit $\rho\rightarrow 1$, with $\rho<1$, we let $\rho=1-\epsilon$, where $\epsilon\downarrow 0$.  Therefore, $\rho^{\pi/2\beta}\sim 1-[\pi/(2\beta)]\epsilon$, hence 
\[
1-\rho^{\pi/(2\beta)}\sim \frac{\pi}{2\beta}(1-\rho),\qquad\text{as }\epsilon\downarrow 0.
\]
Similarly,
\[
1+\rho^{\pi/(2\beta)}\sim 2,\qquad\text{as }\epsilon\downarrow 0.
\]
Hence,
\begin{equation}
G(\rho,0)\sim \frac{\beta}{\pi}\ln \left[\frac{2}{\frac{\pi}{2\beta}(1-\rho)}\right],\qquad\text{as }\epsilon\downarrow 0,
\label{eq:GsimVIP}
\end{equation}
or:
\[
G(\rho,0)\sim -\frac{\beta}{\pi}\ln (1-\rho),\qquad\text{as }\rho\uparrow 1.
\]
The proof for {\textbf{Case 2}}, with $\varphi=0$ and $\rho>1$ is exactly the same, and again leads to the required behaviour:
\[
G(\rho,0)\sim -\frac{\beta}{\pi}\ln (\rho-1),\qquad\text{as }\rho\downarrow 1.
\]
So overall,
\[
G(\rho,0)\sim -\frac{\beta}{\pi}\ln |1-\rho|,\qquad\text{as }\rho\rightarrow 1.
\]
as required. \myqed
\end{proof}
To match the outer solution, we will take:
\[
v=-\frac{\cancel{\pi}}{\beta}\frac{Q(t)}{\cancel{\pi} D_v} G(\rho,\varphi),
\]
hence:
\begin{equation}
c_{inner}(\rho,\varphi,t)=c_s-\frac{Q(t)}{ D_v\beta}\sum_{n=0}^\infty \frac{1}{\lambda_n}\cos(\lambda_n\varphi)F_n(\rho).
\label{eq:cInnerFinal}
\end{equation}
The Laplace-transformed version of $c_{inner}$ is therefore:
\begin{equation}
\widehat{c}_{inner}(\rho,\varphi,s)=\frac{c_s}{s}-\frac{\widehat{Q}(s)}{ D_v\beta}\sum_{n=0}^\infty \frac{1}{\lambda_n}\cos(\lambda_n\varphi)F_n(\rho).
\label{eq:cInnerFinalL}
\end{equation}
where $\widehat{Q}(s)$ is the Laplace transform of $Q(t)$.

\section{Matching inner and outer solutions}

In this section, we match the inner and outer solutions of the diffusion equation~\eqref{eq:model_all} across an overlap region.  The inner solution is developed in the main manuscript and in Section~\ref{sec:inner} above.  Here, to make the supplementary material more self-contained, we first recall the outer solution.

\subsection{Outer solution}

In the far field, for $r\gg a$, we may work in the physical ($xy$) space since the precise droplet geometry is not important.   For the same reason, radial symmetry is assumed.  Consequently, the diffusion equation~\eqref{eq:model_all} reduces to:
\begin{equation}
\frac{\partial c}{\partial t}=D_v\frac{\partial }{\partial r}\left(r\frac{\partial c}{\partial r}\right).
\label{eq:diffusionRad}
\end{equation}
We introduce the Laplace transform of $c$:
\begin{equation}
\hatcL(r,s)=\int_0^\infty c(r,t)\mathe^{-st}\mathd t.
\end{equation}
The diffusion equation~\eqref{eq:diffusionRad} may itself be Laplace-transformed to give:
\begin{equation}
c_\infty+s\hatcL=D_v\frac{1}{r}\frac{\partial}{\partial r}\left(r\frac{\partial\hatcL}{\partial r}\right).
\label{eq:diffusionL}
\end{equation}
The solution to Equation~\eqref{eq:diffusionL} is:
\begin{equation}
\hatcL(r,s)=\frac{c_\infty}{s}+A(s)K_0\left( \frac{s^{1/2}r}{D_v^{1/2}}\right),
\end{equation}
where $K_0$ is the modified Bessel function of the second kind.  We use the small-argument approximation of $K_0$ to obtain:
\begin{equation}
\hatcL(r,s)\sim \frac{c_\infty}{s} +A(s)\left[\ln \left(\frac{2D_v^{1/2}}{rs^{1/2}}\right)-\gamma_E\right],
\label{eq:hatcLSln}
\end{equation}
valid as $r\rightarrow 0$.  Here, $\gamma_E$ is the Euler--Mascheroni constant.  In the mapped space, Equation~\eqref{eq:hatcLSln} can be written as:
\begin{equation}
\widehat{c}_{outer}(s,r)= \frac{c_\infty}{s} +A(s)\left[ \ln|\zeta-\zeta_0|+\ln\left(\frac{2 (D_v/s)^{1/2}}{|z_+-z_-|}\right)-\gamma_E\right],
\label{eq:cOuterLaplaceMatch}
\end{equation}
where $\zeta_0=-\mathe^{\imag\alpha}$.  In equation~\eqref{eq:cOuterLaplaceMatch}, we have added the  subscript `outer' has been added to $\widehat{c}$, emphasizing that this particular form will be used to match the outer solution for $c$ with the inner solution.

\subsection{Matching}

Going back to the proof of Theorem~\ref{thm:Glog} (see especially, Equation~\eqref{eq:GsimVIP}), on the boundary faces, in the limit of large $|x|$, we have:
\[
\widehat{c}_{inner}(r,s)\sim \frac{c_s}{s}-\frac{\widehat{Q}(s)}{ D_v\beta} \bigg\{ \frac{\beta}{\pi}\left[-\ln|1-\rho|+\ln(4\beta/\pi)\right]\bigg\},
\]
hence
\[
\widehat{c}_{inner}(r,s)\sim \frac{c_s}{s}+\frac{\widehat{Q}(s)}{\pi D_v} \left[ \ln|1-\rho|-\ln(4\beta/\pi)\right],
\]
Referring back to Equation~\eqref{eq:cOuterLaplaceMatch}, in the overlap region, and on boundary faces, we have:
\[
\widehat{c}_{outer}= \frac{c_\infty}{s} +A(s)\left[ \ln|\zeta-\zeta_0| +\ln \left(\frac{2(D_v /s)^{1/2}}{|z_+-z_-|}\right)-\gamma_E+\cdots\right],\qquad \text{as }\zeta\rightarrow \zeta_0.
\]
Since $\ln|\zeta-\zeta_0|=\ln|1-\rho|$ we  have, matching other terms in the inner and outer solution in the overlap region,
\[
\frac{c_s}{s}+\frac{\widehat{Q}(s)}{\pi D_v}\left[\cancel{\ln|1-\rho|}-\ln(4\beta/\pi)\right]=\frac{c_\infty}{s}+A(s)\left[\cancel{\ln|\zeta-\zeta_0|}+\ln \left(\frac{2(D_v /s)^{1/2}}{|z_+-z_-|}\right)-\gamma_E\right].
\]
where the cancellation occurs provided we take:
\[
A(s)=\frac{\widehat{Q}(s)}{\pi D_v}.
\]
This gives a closed expression for $\widehat{Q}(s)$:
\[
\widehat{Q}(s)=\frac{1}{2}\frac{(c_s-c_\infty)\pi D_v}{\ln  \left(\frac{2(D_v /s)^{1/2}}{|z_+-z_-|}\right)+\ln(4\beta/\pi)-\gamma_E}.
\]
Alternatively, we may write this as:
\[
\widehat{Q}(s)=\frac{1}{2}\frac{(c_s-c_\infty)\pi D_v}{\tfrac{1}{2}\ln  \left(\frac{16(D_v /s)}{a^2}\right)+\ln(\beta/\pi)-\gamma_E}.
\]
This result (with non-zero $\alpha$) is the same in form as in Yariv's paper~\cite{yariv2023lifetime}, provided we take:
\[
\gamma_{eff}=\gamma_E-\ln(\beta/\pi).
\]
Consequently, we may jump to an expression for $Q(t)$ in real space:
\begin{equation}
Q(t)=\tfrac{1}{2}(c_s-c_\infty)\pi D_v\int_0^\infty \mathe^{-\lambda t} \frac{1}{\left[\tfrac{1}{2}\ln  \left(\frac{16(D_v /s)}{a^2}\right)-\gamma_{eff}\right]^2}\frac{\mathd \lambda}{\lambda}.
\label{eq:yarinF}
\end{equation}

\subsection{Expression for the evaporative flux}

From Equation~\eqref{eq:Jdef}, we may work out the evaporative flux for a 2D cylindrical droplet, using the inner solution to compute the vapour-concentration gradient.  Taking account of the metric factor induced by transforming between $xy$ coordinates and $\xi\eta$ coordinates, the evaporative flux is:
\begin{equation}
J=-D_v(\hatn\cdot\nabla_{xy}c)_{r=R}=D_v\left|\frac{\mathd f}{\mathd z}\right|_{\eta=0}\left(\frac{\partial c}{\partial \eta}\right)_{\eta=0}.
\label{eq:J2Dexpr1}
\end{equation}
Using the property of the metric factor $|\mathd f/\mathd z|$ in Equation~\eqref{eq:dfdzMetric} this becomes:
\begin{equation}
J=D_v\frac{\rho^2+2\rho\cos\alpha+1}{2a}\left(\frac{\partial c}{\partial \eta}\right)_{\eta=0}.
\label{eq:J2Dexpr2}
\end{equation}
The inner solution~\eqref{eq:cInnerFinal} is substituted into~\eqref{eq:J2Dexpr2} and evaluated at $\eta=0$ ($\varphi=\beta$) to produce a final closed-form expression for the evaporative flux, as a function of the polar angle $\theta$ in the physical space:
\begin{subequations}
\begin{equation}
J(\theta)=\frac{Q(t)}{2a\beta}\left(\rho^2+2\rho\cos\alpha+1\right)\frac{1}{\rho}\frac{\rho^{\pi/2\beta}}{1+\rho^{\pi/\beta}},
\end{equation}
where $\rho$ is constrained:
\begin{equation}
\rho=\mathcal{R}(\theta),\qquad \mathcal{R}(\theta)=-\frac{\sin\left(\frac{\theta-\theta_+}{2}\right)}{\sin\left(\frac{\theta-\theta_-}{2}\right)},\qquad \theta_+<\theta<\theta_-.
\end{equation}
\label{eq:Jfinal}
\end{subequations}

\section{Evaluation of integral flux in the large-$t$ limit}

We may evaluate $Q(t)$ in the late-time limit.  The starting-point for the calculation is the large-$t$ evaluation of the integral
\[
\myInt=\int_0^\infty \mathe^{-\lambda t}\frac{1}{ \left[\tfrac{1}{2}\ln \left(\frac{16D_v}{\lambda a^2}\right)-\gamma_{eff}\right]^2+\tfrac{1}{4}\pi^2}\frac{\mathd\lambda}{\lambda}.
\]
We have:
\begin{eqnarray*}
\myInt
 &=&4\int_0^\infty \mathe^{-\lambda t}\frac{1}{ \left[\ln \left(\frac{16D_v}{\lambda a^2}\right)-2\gamma_{eff}\right]^2+\pi^2}\frac{\mathd\lambda}{\lambda},\\
  &=&4\int_0^\infty \mathe^{-\lambda t}\frac{1}{ \left[\ln \left(\frac{16D_v}{\lambda a^2}\mathe^{- 2\gamma_{eff}}\right)\right]^2+\pi^2}\frac{\mathd\lambda}{\lambda}.
\end{eqnarray*}
Let:
\[
\mu =\frac{\lambda a^2}{16D_v}\mathe^{2\gamma_{eff}}.
\]
Also let:
\[
\tau=\frac{16 D_v}{a^2}\mathe^{-2\gamma_{eff}}t.
\]
Therefore: 
\[
\frac{\mathd\lambda}{\lambda}=\frac{\mathd \mu}{\mu},\qquad \mu \tau=\lambda t.
\]
So we have:
\[
\myInt=4\int_0^\infty \mathe^{-\mu \tau}\frac{1}{ (\ln \mu)^2+\pi^2}\frac{\mathd\mu}{\mu}.
\]
We make one more substitution: $u=\mu \tau$, so we have:
\[
\myInt=4\underbrace{\int_0^\infty \mathe^{-u}\frac{1}{(\ln u-\ln \tau)^2+\pi^2}\frac{\mathd u}{u}}_{=\myJint(\tau)}.
\]

\subsection{Evaluation of $\myJint(\tau)$}

We have:
\begin{equation}
\frac{1}{A^2+\pi^2}=\frac{1}{\pi}\int_0^\infty \mathe^{-As}\sin(\pi s)\mathd s.
\label{eq:lapltrA}
\end{equation}
Also (A\&S 6.5.3):
\[
\int_0^t v^{a-1}\mathe^{-v}\mathd v=\gamma(a,t)=\Gamma(a)-\Gamma(a,t).
\]
Sub into Equation~\eqref{eq:lapltrA}:
\begin{equation}
\frac{1}{(\ln t-\ln v)^2+\pi^2}=\frac{1}{\pi}\int_0^\infty \mathe^{-s(\ln t-\ln v)}\sin(\pi s)\mathd s,
\end{equation}
hence:
\begin{equation}
\frac{1}{(\ln t-\ln v)^2+\pi^2} =\frac{1}{\pi}\int_0^\infty t^{-s}v^s \sin(\pi s)\mathd s.
\label{eq:lapltrA2}
\end{equation}

As $\tau\rightarrow\infty$, the integral $\myJint(\tau)$ is equivalent to:
\[
\myJint(\tau)=\int_0^\tau \mathe^{-u}\frac{1}{(\ln u-\ln \tau)^2+\pi^2}\frac{\mathd u}{u}.
\]
We therefore have:
\begin{eqnarray*}
\myJint(\tau)&=&\int_0^\tau \left[\frac{1}{(\ln t-\ln v)^2+\pi^2} \right]\frac{\mathe^{-v}}{v},\\
    &\stackrel{\text{Eq.~\eqref{eq:lapltrA2}}}{=}&\frac{1}{\pi}\int_0^\tau  \left[\int_0^\infty \tau^{-s}v^s \sin(\pi s)\mathd s\right]\frac{\mathe^{-v}}{v}\mathd v,\\
		&=&\frac{1}{\pi}\int_0^\infty \tau^{-s}\sin(\pi s)\left[\int_0^\tau v^{s-1}\mathe^{-v}\mathd v\right]\mathd s.
\end{eqnarray*}
Let:
\[
L=\ln (\tau) \implies \tau^{-s}=\mathe^{-s\ln \tau}=\mathe ^{-sL}.
\]
Hence:
\[
\myJint=\frac{1}{\pi}\int_0^\infty \mathe^{-sL}\sin (\pi s) \underbrace{\left[\int_0^\tau v^{s-1}\mathe^{-v}\mathd v\right]\mathd s}_{=\gamma(s,\tau)}.
\]
Here, we are using the lower incomplete gamma function $\gamma(s,\tau)$.
We split the integral $J$ into two parts:
\[
\myJint=\myJint_1+\myJint_2, 
\]
where:
\begin{eqnarray*}
\myJint_1&=&\frac{1}{\pi}\int_0^\delta \mathe^{-sL}\sin (\pi s) \gamma(s,\tau)\mathd s,\\
\myJint_2&=&\frac{1}{\pi}\int_\delta^\infty \mathe^{-sL}\sin (\pi s) \gamma(s,\tau)\mathd s.
\end{eqnarray*}
Here, $\delta$ is a small positive arbitrary number.  We take the second integral first.   We have:
\begin{eqnarray*}
\mathe^{-sL}\gamma(s,t)&=&\tau^{-s}\int_0^\tau v^{s-1}\mathe^{-v}\mathd v,\\
                       &=&\int_0^\tau (v/\tau)^s\mathe^{-v}\frac{\mathd v}{v},\\
											&\stackrel{v=\tau x}{=}&\underbrace{\int_0^1 x^{s-1}\mathe^{-\tau x}\mathd x}_{=a_\tau(s)}.
\end{eqnarray*}
We have:
\[
a_\tau(s)=\int_0^1 x^{s-1}\mathe^{-\tau x}\mathd x=\tau^s \gamma(\delta,\tau).
\]
Notice:
\[
\lim_{s\rightarrow\infty}a_\tau(s)=0,
\]
and
\[
a_\tau(s)>0.
\]
Also,
\[
\frac{\mathd a_\tau}{\mathd s}=\int_0^1 x^{-1}x^s \ln(x)\mathe^{-\tau x}\mathd x.
\]
As $x<1$ on the domain of integration, we have $\ln(x)<0$, hence $a_\tau'(s)<0$.  We now use integration by parts on $\myJint_2$ to get:
\begin{eqnarray*}
\myJint_2&=&\frac{1}{\pi}\int_\delta^\infty \sin(\pi s)a_\tau(s)\mathd s,\\
&=&-\frac{a_\tau(s)\cos(\pi s)}{\pi^2}\big|_\delta^\infty+\frac{1}{\pi^2}\int_\delta^\infty \cos(\pi s)a_\tau
\end{eqnarray*}
Therefore:
\begin{eqnarray*}
|\myJint_2|&\leq& \frac{a_\tau(\delta)}{\pi^2}+\frac{1}{\pi^2}\int_\delta^\infty |a_\tau'(s)|\mathd s,\\
     &\leq& \frac{a_\tau(\delta)}{\pi^2}+\frac{1}{\pi^2}\int_\delta^\infty \left[-a_\tau'(s)\right]\mathd s,\\
		&\leq & \frac{a_\tau(\delta)}{\pi^2}+\frac{1}{\pi^2}\lim_{\ell\rightarrow\infty}[-a_\tau'(s)]_\delta^\ell,\\
		&=&\frac{2a_\tau(\delta)}{\pi^2}.
\end{eqnarray*}
But:
\[
a_\tau(\delta)=\tau^{-\delta}\gamma(\delta,\tau)\leq \tau^{-\delta}\Gamma(\delta).
\]
Therefore, for fixed $\delta$,
\[
|\myJint_2|\leq \frac{2}{\pi^2}\tau^{-\delta}\Gamma(\delta),
\]
which vanishes comfortably as $\tau\rightarrow\infty$.

For the integral $\myJint_1$, we use:
\[
\gamma(s,\tau)\sim \Gamma(s)\qquad\text{as }\tau\rightarrow\infty.
\]
Therefore:
\[
\myJint_1\sim \frac{1}{\pi}\int_0^\delta \mathe^{-sL}\sin (\pi s) \Gamma(s)\mathd s,\qquad \text{as }\tau\rightarrow\infty.
\]
We use the expansions:
\begin{eqnarray*}
\sin(\pi s)&=& \pi s-\tfrac{1}{6}\pi^3 s^3+\cdots,\\
\Gamma(s)&=&\frac{1}{s}-\gamma_E+\tfrac{1}{2}\left(\gamma_E^2+\frac{\pi^2}{6}\right)s+\cdots
\end{eqnarray*}
Hence:
\begin{eqnarray*}
\frac{1}{\pi}\sin(\pi s)\Gamma(s)&=&\left(s-\tfrac{1}{6}\pi^2 s^3\right)\left[\frac{1}{s}-\gamma_E+\tfrac{1}{2}\left(\gamma_E^2+\frac{\pi^2}{6}\right)s\right]+\cdots,\\
&=&1-\gamma_E s +\tfrac{1}{2}\left(\gamma_E^2-\tfrac{1}{6}\pi^2\right)s^2+\cdots
\end{eqnarray*}
Therefore:
\[
\myJint_1\sim \int_0^\delta \mathe^{-sL}\left(1-\gamma_E s+\cdots\right)\mathd s\qquad\text{as }\tau\rightarrow\infty.
\]
The remaining integrals can be done analytically and we get:
\[
\myJint_1\sim \frac{1}{L}\left(1-\mathe^{-\delta L}\right)-\frac{1}{L^2}\left[\gamma_E\left(1-\mathe^{-\delta L}\right)-\frac{\delta}{L}\mathe^{-\delta L}\right]+\cdots,
\]
hence
\[
\myJint\sim \frac{1}{\ln \tau}-\frac{\gamma_E}{(\ln \tau)^2}+\cdots\qquad \text{as }\tau\rightarrow \infty.
\]
This is equivalent to:
\[
\myJint\sim \frac{1}{\ln \tau +\gamma_E}+\cdots\qquad \text{as }\tau\rightarrow \infty.
\]
Therefore:
\begin{eqnarray*}
\myJint&\sim& \frac{1}{\ln \left(\frac{16 D_v}{a^2}\mathe^{-2\gamma_{eff}} t\right)+\gamma_E},\\
  &=&      \frac{1}{\ln \left( \frac{D_v}{a^2} t\right)+\ln(16)-2\gamma_{eff}+\gamma_E},\\
  &=&				 \frac{1}{\ln \left( \frac{D_v}{a^2} t\right)+\ln(16)-2\gamma_E+2\ln(\beta/\alpha)+\gamma_E},\\
  &=& \frac{1} {2\ln \left(\frac{\sqrt{D_v t}}{a}\right)+2\ln(4)-\gamma_E+2\ln(\beta/\alpha)},\\
	&=& \frac{1} {2\ln \left(\frac{\sqrt{D_v t}}{a}\right)+2\ln(4)-\gamma_E+2\ln(\beta/\alpha)},\\
	&=& \frac{1} {2\ln \left(\frac{\sqrt{D_v t}}{a}\right)-\gamma_E+2\ln(4\beta/\alpha)},\\
		&=&\tfrac{1}{2} \frac{1} {\ln \left(\frac{\sqrt{D_v t}}{a}\right)+\ln(4\beta/\alpha)-\tfrac{1}{2}\gamma_E},\\
\end{eqnarray*}
Reasoning back up the chain, we get:
\begin{equation}
Q(t)\sim  \frac{\pi D_v(c_s-c_\infty)}{\ln \left(\frac{\sqrt{D_v t}}{a}\right)+\ln(4\beta/\alpha)-\tfrac{1}{2}\gamma_E}.
\label{eq:QtAsy}
\end{equation}

\section{The Stokes-Flow Problem in the Mapped Domain}

We now look at the flow inside the droplet generated by the evaporative flux.  Let the flow be denoted by $\vecu$, with Cartesian components $u$ and $v$.  Rather than starting with a full formulation of the equations satisfied by $\vecu$, we will introduce constraints on $\vecu$ progressively, starting with the simplest ones.

\subsection{Incompressibility}

The first constraint is that the flow is incompressible: $\nabla\cdot\vecu=0$.  Therefore, a streamfunction $\psi$ exists, such that:
\[
u=\frac{\partial\psi}{\partial  y},\qquad v=-\frac{\partial\psi}{\partial x}.
\]
Alternatively, we may write:
\begin{equation}
\vecu=\hatz\times\nabla\psi.
\label{eq:hatzPsi}
\end{equation}

\subsection{Mass conservation}

We look at the mass-conservation law for fluid mass in an evaporating droplet.  Let the area of the droplet be $A(t)$.  Also, let $\Omega(t)$ denote the domain occupied by the droplet.    Hence:
\begin{equation}
A(t)=\int_{\Omega(t)}1\mathd^2 x,
\end{equation}
We denote the boundary of the domain $\Omega(t)$ as $\partial\Omega(t)$.  This boundary is made up of a flat substrate $F$ and a curved surface $S$.  The surface $S$ was introduced previously in the context of the atmospheric diffusion equation, although now we  accept the possibility that $S$ may be time-dependent.   We assume that we are in the pinned contact-line mode (CR), so $F$ is a time-independent domain.   Hence:
\begin{equation}
\partial\Omega(t)=S(t)\cup F.
\end{equation}
Using Reynolds Transport Theorem, we get:
\begin{equation}
\frac{\mathd A}{\mathd t}=\int_{S(t)}V_n \mathd \ell +\int_{\Omega(t)}(-\nabla\cdot\vecu)\mathd^2 x,
\label{eq:dVdt1}
\end{equation}
where $V_n=\partial R/\partial t$ is the normal velocity of the interface.  Equivalently,
\begin{equation}
\frac{\mathd A}{\mathd t}=-\int_{S(t)}(J/\rho_\ell)\mathd \ell,
\label{eq:dVdt2}
\end{equation}
where $J$ is the flux of mass leaving the droplet and $\rho_\ell$ is the liquid density.    Equating the 
expressions~\eqref{eq:dVdt1} and~\eqref{eq:dVdt2} gives:
\[
\int_{S(t)}V_n \mathd \ell +\int_{\Omega(t)}(-\nabla\cdot\vecu)\mathd^n x=-\int_{S(t)}(J/\rho_\ell)\mathd \ell.
\]
Applying the Divergence Theorem gives:
\[
V_n-\hatn\cdot\vecu=-\frac{J}{\rho_{\ell}},\qquad \text{on }S(t).
\]
Here, $\hatn$ is the outward-pointing unit normal to the boundary $S(t)$.    
%
Re-arranging gives:
\begin{equation}
V_n+\frac{J}{\rho_\ell}=\hatn\cdot\vecu.
\label{eq:VnBc}
\end{equation}
Since $\nabla\cdot\vecu=0$ inside the fluid domain (for an incompressible fluid), Equations~\eqref{eq:dVdt1} and~\eqref{eq:dVdt2} also provide the following compatibility condition:
\[
\int_{S(t)}V_n \mathd \ell +\int_{S(t)}\frac{J}{\rho_{\ell}}\mathd \ell=0.
\]
%

\subsection{Expression for $V_n$}

We now formulate an expression for $V_n$.  Care must be taken here, as we wish to work in a Galilean frame in which the flat faces $F_-$, $F$, and $F_+$ are stationary (lab frame).  We therefore define $y=Y+d$, such that $Y=0$ is the substrate location.  Therefore, relative to the lab frame, the origin of the $xy$ coordinate system moves with velocity
\begin{equation}
\bm{U}_c=(0,-\dot d),
\end{equation}
Therefore, relative to the lab frame, we have:
\begin{equation}
V_n(\theta,t)=-\dot d\sin\theta+\dot R.
\label{eq:Vn}
\end{equation}

To obtain an ordinary differential equation for the droplet size,  we use the expression $M=\rho_\ell a^2\mathcal{A}(\alpha)$ for the droplet mass.  Here, $a^2\mathcal{A}(\alpha)$ is the droplet area, where
\begin{equation}
\mathcal{A}(\alpha)=\frac{2\alpha-\sin(2\alpha)}{2\sin^2\alpha},\qquad \alpha \neq 0,\pi.
\end{equation}
Then, by conservation of mass,
\begin{equation}
\frac{\mathd}{\mathd t}\left[\rho_\ell a^2\mathcal{A}(\alpha)\right]=\underbrace{\int_{\theta_-}^{\theta_+}J R\,\mathd \theta}_{=F_{2D}}.
\label{eq:massConsIntegral}
\end{equation}
The quantity $F_{2D}$ is the integral flux in two dimensions.  
To make further progress, we must specify the mode of evaporation as either constant radius (CR: $a=\text{Const.}$) or constant contact angle (CA: $\alpha=\text{Const.}$).  Here, we look at both modes of evaporation separately.

\paragraph*{CR Mode:}

We have $a=\text{Const}$.  With $R^2=a^2+d^2$ and $d/R=\cos\alpha$, this gives $R\dot R=d\dot d$, and hence $\dot R=(d/R)\dot d=\cos\alpha \dot d$, so Equation~\eqref{eq:Vn} reduces to:
\begin{equation}
V_n(\theta,t)=\dot d \left(\cos\alpha-\sin\theta\right).
\label{eq:VnCR1}
\end{equation}
From differentiating $(\mathd /\mathd t)(d\tan\alpha)=0$ we get $\dot d=-a\dot\alpha/\sin^2\alpha$, hence:
\begin{equation}
V_n(\theta,t)=-\frac{a\dot\alpha}{\sin^2\alpha} \left(\cos\alpha-\sin\theta\right).
\label{eq:VnCR2}
\end{equation}
Note that $V_n(\pi/2\pm \alpha,t)=0$, which is consistent with the CR assumption.  
From Equation~\eqref{eq:massConsIntegral} we also have:
\begin{equation}
\rho_\ell a^2 \frac{\mathd\mathcal{A}}{\mathd \alpha}\dot\alpha=F_{2D},
\label{eq:massConsIntegral1}
\end{equation}
which is a closed-form ODE for $\alpha$.  The expression for $\dot\alpha$ may then be used to close the expression~\eqref{eq:VnCR2} for $V_n$.

\paragraph*{CCA Mode:}

We have $R\dot R=a\dot a+d\dot d$, hence $\dot R=(a/R)\dot a+(d/R)\dot d$, and $\dot R=\sin\alpha \dot a+\cos\alpha \dot d$.  But $d\tan\alpha=a$ and $\alpha=\text{Const}$ so $\dot d\tan\alpha= \dot a$, hence:
\begin{eqnarray*}
\dot R&=&\sin \alpha \dot d\tan\alpha+\cos\alpha \dot d,\\
      &=&\dot d\left(\sin\alpha\tan\alpha+\cos\alpha\right),\\
			&=&\frac{\dot d}{\cos\alpha},
\end{eqnarray*}
hence
\begin{equation}
\dot R\cos\alpha=\dot d.
\end{equation}
Hence, Equation~\eqref{eq:Vn} reduces to:
\begin{equation}
V_n(\theta,t)=-\dot R \left(\cos\alpha\sin\theta-1\right).
\label{eq:VnCCA1}
\end{equation}
We use $\dot d \tan\alpha=\dot a$ to re-write $\dot R$ as:
\begin{equation}
\dot R=\frac{\dot a}{\sin\alpha}.
\end{equation}
Hence, Equation~\eqref{eq:VnCCA1} can be re-written as:
\begin{equation}
V_n(\theta,t)=-\frac{\dot a}{\sin\alpha} \left(\cos\alpha\sin\theta-1\right).
\label{eq:VnCCA2}
\end{equation}
Note that $V_n(\pi/2\pm \alpha,t)=\dot a \sin\alpha$, consistent with the CCA assumption.
From Equation~\eqref{eq:massConsIntegral} we also have:
\begin{equation}
2\rho_\ell \mathcal{A}a\dot a=F_{2D},
\label{eq:massConsIntegral2}
\end{equation}
which is a closed-form ODE for $a$.   The expression for $\dot a$ may then be used to close the expression~\eqref{eq:VnCCA2} for $V_n$.

\subsection{Geometry of the conformal map}

We  apply the M\"obius map~\eqref{eq:mobius} to the liquid domain.  Under the M\"obius map, the droplet occupies the wedge $0<\arg(\zeta)<\alpha$, such that the atmosphere and the droplet occupy a half-space $\arg(\zeta)\in (\pi+\alpha,2\pi)\cup (0,\alpha)$, with overlap along the ray $\arg(\zeta)=0$, corresponding to the surface $S$.  

We may further view the coordinates $(\xi,\eta)$ produced by M\"obius map as a special kind of orthogonal curvilinear coordinates.  The Euclidean metric is:
\[
\mathd s^2=\mathd x^2+\mathd y^2.
\]
Under the conformal map:
\[
\mathd s^2=\frac{1}{|\mathd f/\mathd z|^2}\left(\mathd\xi^2+\mathd \eta^2\right).
\]
Using the $(\rho,\phi)$ polar coordinates in the mapped space, we have:
\[
\mathd s^2=\frac{1}{|\mathd f/\mathd z|^2}\left(\mathd\rho^2+\rho^2\mathd\phi^2\right).
\]
Therefore, we may view the $(\rho,\phi)$ polar coordinates as orthogonal curvilinear coordinates with scale factors:
\[
h_\rho=\frac{1}{|\mathd f/\mathd z|},\qquad h_\phi=\frac{\rho}{|\mathd f/\mathd z|}.
\]
In terms of the $(\rho,\phi)$ polar coordinates, the fluid velocity vector is:
\[
\vecu=u_\rho \hatrho+u_\phi\hatphi,
\]
where
\[
u_\rho=\hatrho\cdot\vecu,\qquad u_\phi=\hatphi\cdot\vecu.
\]
Here, the unit vectors are given by:
\begin{subequations}
\begin{eqnarray}
\hatrho&=&\cos\phi\,\hati_{\xi\eta}+\sin\phi\,\hatj_{\xi\eta},\\
\hatphi&=&-\sin\phi\,\hati_{\xi\eta}+\cos\phi\,\hatj_{\xi\eta}.
\end{eqnarray}%
\label{eq:unitRhoPhi}%
\end{subequations}%
Here, the subscript $\xi\eta$ is used to emphasize that these are Cartesian coordinates in the mapped space.
Using Equation~\eqref{eq:hatzPsi}, we have:
\begin{eqnarray*}
u_\rho&=&\hatrho\cdot(\hatz\times\nabla\psi),\\
      &\stackrel{\text{Tr. Prod}}{=}&\nabla\psi\cdot(\hatrho\times\hatz),\\
			&=&\hatphi\cdot\nabla\psi,\\
			&=&\frac{1}{h_\phi}\frac{\partial\psi}{\partial\phi}.
\end{eqnarray*}
Similarly:
\begin{equation}
u_\phi=-\frac{1}{h_\rho}\frac{\partial\psi}{\partial \rho}.
\end{equation}
Hence:
\begin{equation}
u_\rho=\frac{1}{h_\phi}\frac{\partial\psi}{\partial\phi},\qquad u_\phi=-\frac{1}{h_\rho}\frac{\partial\psi}{\partial \rho}
\label{eq:psiRhoPhi}
\end{equation}

\subsection{Streamfunction boundary condition and mapping}

Using the previous geometrical arguments, we compute $\vecu\cdot\hatn$ on $S$ in terms of the coordinates $(\rho,\phi)$.  We have:
\begin{eqnarray*}
\vecu\cdot\hatn&=&(u_\rho\hatrho+u_\phi\hatphi)\cdot\hatn,\\
               &=&(u_\rho\hatrho+u_\phi\hatphi)\cdot\left[ (\hatn\cdot\hatrho)\hatrho+(\hatn\cdot\hatphi)\hatphi\right],\\
							 &\stackrel{\phi=0}{=}&(u_\rho\hatrho+u_\phi\hatphi) \cdot(-\hatphi),\\
               &=& -u_\phi,\\
							 &=&\frac{1}{h_\rho}\frac{\partial\psi}{\partial \rho},\\
							 &=&\left|\frac{\mathd f}{\mathd z}\right|\frac{\partial\psi}{\partial \rho}.
\end{eqnarray*}
Putting all these elements together, Equation~\eqref{eq:VnBc} becomes.
\begin{equation}
\left|\frac{\mathd f}{\mathd z}\right|\frac{\partial\psi}{\partial \rho}=\dot R\left(1-\frac{\sin\theta}{\cos\alpha}\right)+\frac{J(\theta)}{\rho_\ell},\qquad \text{on }S.
\label{eq:bcPsiS}
\end{equation}
We therefore have:
\begin{eqnarray*}
\left|\frac{\mathd f}{\mathd z}\right|\frac{\partial\psi}{\partial \rho}&=&\frac{J(\theta)}{\rho_\ell}+\dot R\left(1-\frac{\sin\theta}{\cos\alpha}\right),\qquad \text{on }S,\\
\frac{\partial\psi}{\partial \rho}&=&\frac{1}{|df/dz|}\frac{J(\theta)}{\rho_\ell}+\cdots,\qquad\text{on }\arg(\zeta)=0,\\
&\stackrel{\text{Eq.~\eqref{eq:Jfinal}}}{=}&\cancel{\frac{1}{|df/dz|}} \frac{Q(t)}{\beta\rho_\ell} \cancel{\frac{\rho^2+1+2\rho\cos\alpha}{2a}}\frac{1}{\rho}\frac{\rho^{\pi/2\beta}}{1+\rho^{\pi/\beta}}+\cdots
\end{eqnarray*}
Therefore:
\begin{equation}
\frac{\partial\psi}{\partial \rho} =  \frac{Q(t)}{\beta\rho_\ell}\frac{\rho^{\frac{\pi}{2\beta}-1}  }{1+\rho^{\pi/\beta}}+\cdots.
\label{eq:dsidRho}
\end{equation}

\subsection{Stokes problem}

To further specify $\psi$ we require it to solve the Stokes problem:
\begin{equation}
\nabla^4\psi=0.
\label{eq:bilaplace}
\end{equation}
This is appropriate for viscous flow satisfying the Stokes equation $-\nabla p+\mu\nabla^2\vecu=0$.  Here, $p$ is the pressure and $\mu$ is the dynamic viscosity.   Equation~\eqref{eq:bilaplace} is written in terms of the physical-space Laplacian $\nabla_{xy}^2=\partial_x^2+\partial_y^2$.  In terms of the mapped-space Laplacian $\nabla_{\xi \eta}^2=\partial_{\xi}^2+\partial_\eta^2$, we have:
\[
\nabla_{\xi\eta}^2 \left(|\mathd f/\mathd z|^2\nabla_{\xi\eta}^2\psi\right)=0,
\]
Since $h_\rho=1/|\mathd f/\mathd z|$ we may re-write this more compactly as:
\[
\nabla_{\xi\eta}^2 \left(h_\rho^{-2}\nabla_{\xi\eta}^2\psi\right)=0.
\]
The remaining boundary conditions on $\vecu$ (equivalently, on $\psi$) are:
\begin{itemize}[noitemsep]
\item No slip on $F$:
\[
\psi_x=0,\qquad \psi_y=0,\qquad \text{ on }F.
\]
\item Stress-free condition on $S$:
\[
\hatt\cdot\bm{\sigma}\cdot\hatn=0,\qquad \text{on } S,
\]
where $\hatt$ is the tangent vector to the curves surface $S$ and 
$\bm{\sigma}$ is the stress tensor.
\end{itemize}

{\textbf{No slip condition:}}  The no-slip conditions   $\psi_x=0$ and $\psi_y=0$ can be written geometrically as $\hatn\cdot\nabla\psi=0$ and $\hatt\cdot\nabla\psi=0$ where $\hatn$ and $\hatt$ are appropriate tangent vectors on $F$.  These are conformally invariant conditions, and map to:
\[
\psi_\xi=0,\qquad \psi_\eta=0,\qquad \text{ on } \widetilde{F},
\]
or, in terms of polar coordinates:
\[
\psi_\rho=0,\qquad \psi_\phi=0,\qquad \text{ on } \arg(\zeta)=\alpha.
\]
This implies that $\psi=\text{Const.}$ on $\arg(\zeta)=\alpha$, so we may take:
\[
\psi=0,\qquad \psi_\rho=0,\qquad \text{ on }\arg(\zeta)=\alpha.
\]

{\textbf{Stress-free condition:}}  The stress-free boundary condition reads:
\[
\hatn\cdot\left[(\hatt\cdot\nabla)\vecu\right]+\hatt\cdot\left[(\hatn\cdot\nabla)\vecu\right]=0,\qquad \text{on }S,
\]
where $\hatt$ is the tangent vector to the surface $S$, and $\hatt\cdot\hatn=0$.  In terms of mapped coordinates, this is:
\[
-\hatphi\cdot\frac{\partial\vecu}{\partial\rho}- \frac{1}{\rho}\hatrho\cdot\frac{\partial\vecu}{\partial\phi}=0,\qquad \vecu=u_\rho\hatrho+u_\phi\hatphi.
\]
We use the properties of $\hatrho$ and $\hatphi$ in Equation~\eqref{eq:unitRhoPhi} to simplify this expression as follows:
\[
\frac{\partial u_\phi}{\partial \rho}+\frac{1}{\rho}\frac{\partial u_\rho}{\partial \phi}-\frac{u_\phi}{\rho}=0.
\]
Using the streamfunction representation~\eqref{eq:psiRhoPhi}, this simplifies to:
\[
\frac{\partial }{\partial \rho}\left(-\frac{1}{h_\rho}\frac{\partial\psi}{\partial\rho}\right)+\frac{1}{\rho}\frac{\partial }{\partial \phi}\left(\frac{1}{h_\phi}\frac{\partial\psi}{\partial\phi}\right)
+\frac{1}{h_\rho}\frac{1}{\rho}\frac{\partial\psi}{\partial\rho}=0,
\]
or:
\[
\frac{\partial }{\partial \rho}\left(-\left|\frac{\mathd f}{\mathd z}\right|\frac{\partial\psi}{\partial\rho}\right)+\frac{1}{\rho^2}\frac{\partial }{\partial \phi}\left(\left|\frac{\mathd f}{\mathd z}\right|\frac{\partial\psi}{\partial\phi}\right)
+\left|\frac{\mathd f}{\mathd z}\right|\frac{1}{\rho}\frac{\partial\psi}{\partial\rho}=0,
\]

\section{Reduction of Stokes flow to $4\times4$ linear problem}

The droplet triple point $z_+$ corresponds to the limit $\rho\rightarrow 0$. Therefore, we may expand the previous relations in the vicinity of $\rho=0$ to derive expressions for the streamfunction and the vorticity.  We look at $h_\rho=1/\left|\mathd f/\mathd z\right|$.  From Equations~\eqref{eq:dfdz1}--\eqref{eq:dfdz2} we have: 
\[
h_\rho=\frac{2a}{\rho^2+2\rho\cos\alpha+1},
\]
hence
\[
h_\rho\sim 2a\qquad \text{as }\rho\rightarrow 0.
\]
Therefore, in the limit of small $\rho$, the metric reduces to a flat one, and we have the following PDE problem:
\begin{subequations}
\begin{equation}
\nabla_{\xi\eta}^4\psi=0,\\
\end{equation}
with boundary conditions:
\begin{equation}
\psi=0,\qquad \psi_\rho=0,\qquad \text{on }\phi=\alpha,
\end{equation}
\begin{equation}
-\frac{\partial^2\psi}{\partial\rho^2}+\frac{1}{\rho^2}\frac{\partial^2\psi}{\partial\phi^2}+\frac{1}{\rho}\frac{\partial\psi}{\partial \rho}=0,\qquad \text{on }\phi=0.
\end{equation}
The remaining boundary condition is supplied by the flux condition~\eqref{eq:dsidRho} particularized to the limit $\rho\rightarrow 0$:
\begin{equation}
\frac{\partial\psi}{\partial \rho} = \frac{Q(t)}{\beta\rho_\ell}\rho^{\frac{\pi}{2\beta}-1}.
\end{equation}
\end{subequations}
The form of this PDE is the same as the `corner-eddy problem' proposed by Moffatt and sketched out in Acheson~\cite{acheson1990elementary}.
For this reason, we propose a solution:
\begin{equation}
\psi=\rho^{p+1}F(\phi),
\end{equation}
where 
\begin{equation}
p=\frac{\pi}{2\beta}-1,\qquad F(\phi)=c_1\cos[(p+1)\phi]+c_2\sin[(p+1)\phi]+c_3\cos[(p-1)\phi]+c_4\sin[(p-1)\phi],
\end{equation}
and where the coefficients $c_1,\cdots,c_4$ are to be determined.  The first two boundary conditions give:
\begin{equation}
F(\alpha)=0,\qquad F'(\alpha)=0.
\end{equation}
The flux boundary condition gives:
\begin{equation}
(p+1)F(0)=\frac{Q(t)}{\beta\rho_\ell}.
\end{equation}
The zero-stress boundary condition reduces to:
\begin{equation}
F''(0)+(1-p^2)F(0)=0,
\end{equation}
or
\begin{equation}
p\left[c_1(p+1)+c_3(p-1)\right]=0.
\end{equation}
So we are reduced to a $4\times 4$ system:
\begin{multline*}
\left(\begin{array}{cccc} \cos[(p+1)\alpha]& \sin[(p+1)\alpha]&\cos[(p-1)\alpha]&\sin[(p-1)\alpha]\\ 
                          -(p+1)\sin[(p+1)\alpha]& (p+1)\cos[(p+1)\alpha]&-(p-1)\sin[(p-1)\alpha]&(p-1)\cos[(p-1)\alpha]\\ 
													p+1 & 0 & p+1 & 0 \\
													p+1 & 0 & p-1 & 0\end{array}\right)
\left(\begin{array}{c} c_1 \\ c_2 \\ c_3 \\ c_4 \end{array}\right)\\
=
\left(\begin{array}{c} 0 \\ 0 \\ \frac{Q(t)}{\beta\rho_\ell} \\ 0 \end{array}\right)
\end{multline*}
The structure of the linear problem is simple, and we obtain, in the non-degenerate cases:
\begin{equation}
c_3=\tfrac{1}{2}\frac{Q(t)}{\beta\rho_\ell},\qquad c_1=-\frac{p-1}{p+1}c_3,\qquad p\neq -1,
\end{equation}
\begin{equation}
c_2=\underbrace{\left(\frac{p-1}{p+1}\right)\left[\frac{p+1-p\cos (2\alpha)+\cos (2p\alpha)}{\sin (2p\alpha)-p\sin (2\alpha)}\right]}_{=g_1(\alpha,p)}c_3,
\end{equation}
and
\begin{equation}
c_4=\underbrace{\left[\frac{p-1-p\cos(2\alpha)-\cos(2p\alpha)}{\sin (2p\alpha)-p\sin (2\alpha)}\right]}_{=g_2(\alpha,p)}c_3.
\end{equation}
Hence:
\begin{equation}
F(\phi)=\frac{Q(t)}{2p\beta\rho_\ell}\bigg\{-\frac{p-1}{p+1}\cos[(p+1)\phi]+g_1(\alpha,p)\sin[(p+1)\phi]+\cos[(p-1)\phi]+g_2(\alpha,p)\sin[(p-1)\phi]\bigg\}.
\end{equation}
We next compute the vorticity $\omega$, in the small-$\rho$ limit.  We have:
\begin{equation}
\omega=-\frac{1}{4a^2}\nabla_{\xi\eta}^2\psi.
\end{equation}
Since terms like $\rho^{p+1}\sin[(p+1)\phi]$ and $\rho^{p+1}\cos[(p+1)\phi]$ are harmonic, we immediately get:
\begin{eqnarray*}
\omega&=&-\frac{1}{4a^2}\frac{Q(t)}{2p\beta\rho_\ell} \nabla_{\xi\eta}^2 \rho^{p+1}\bigg\{\cos[(p-1)\phi]+g_2(\alpha,p)\sin[(p-1)\phi]\bigg\},\\
      &=&-\frac{1}{4a^2}\frac{Q(t)}{2p\beta\rho_\ell} 4p\rho^{p-1}\bigg\{\cos[(p-1)\phi]+g_2(\alpha,p)\sin[(p-1)\phi]\bigg\},\\
			&=&-\frac{1}{2a^2}\frac{Q(t)}{\beta\rho_\ell} \rho^{p-1}\bigg\{\cos[(p-1)\phi]+g_2(\alpha,p)\sin[(p-1)\phi]\bigg\}.
\end{eqnarray*}

\newpage


\end{document}